\documentclass[]{pasj01}
\usepackage{bm}
\usepackage{lineno}
\Received{}%{yyyy/mm/dd}
\Accepted{}%{yyyy/mm/dd}
\begin{document} 

\title{ X-ray Line Diagnostics of Ion Temperature at Cosmic-Ray Accelerating Collisionless Shocks}

%%% begin:list of authors
% Do NOT capitalize all letters in "textsc".
%\author{A-Firstname \textsc{A-Familyname}\altaffilmark{1}%
%\thanks{Example: Present Address is xxxxxxxxxx}}
%\altaffiltext{1}{A-Address of Institute}
%\email{aaaaa@xxx.xxx.xx.xx}
%
%\author{B-Firstname \textsc{B-Familyname},\altaffilmark{2}}
%\altaffiltext{2}{B-Address of Institute}
%\email{bbbbb@xxx.xxx.xx.xx}
%
%\author{C-Firstname \textsc{C-Familyname}\altaffilmark{3}}
%\altaffiltext{3}{C-Address of Institute}
%\email{ccccc@xxx.xxx.xx.xx}

\author{Jiro \textsc{shimoda}$^*$\altaffilmark{1}}
\altaffiltext{1}{Department of Physics, Graduate School of Science, Nagoya University, \\
Furo-cho, Chikusa-ku, Nagoya 464-8602, Japan}
\email{shimoda.jiro@k.mbox.nagoya-u.ac.jp}

\author{Yutaka \textsc{ohira}\altaffilmark{2}}
\altaffiltext{2}{Department of Earth and Planetary Science, The University of Tokyo, 7-3-1 Hongo, Bunkyo-ku, Tokyo 113-0033, Japan}

\author{Aya \textsc{bamba}\altaffilmark{3,4}}
\altaffiltext{3}{Department of Physics, Graduate School of Science, the University
of Tokyo, 7-3-1 Hongo, Bunkyo-ku, Tokyo 113-0033, Japan}
\altaffiltext{4}{Research Center for the Early Universe, School of Science, The
University of Tokyo, 7-3-1 Hongo, Bunkyo-ku, Tokyo 113-0033, Japan}

\author{Yukikatsu \textsc{Terada}\altaffilmark{5,6}}
\altaffiltext{5}{Graduate School of Science and Engineering, Saitama University, 255 Shimo-Ohkubo, Saitama, 338-8570, Japan}
\altaffiltext{6}{Institute of Space and Astronautical Science, Japan Aerospace Exploration Agency,
3-1-1 Yoshinodai, Chuo, Sagamihara, Kanagawa, 252-5210, Japan}

\author{Ryo \textsc{Yamazaki}\altaffilmark{7,8}}
\altaffiltext{7}{Department of Physical Sciences, Aoyama Gakuin University, 5-10-1 \\
Fuchinobe, Sagamihara 252-5258, Japan}
\altaffiltext{8}{Institute of Laser Engineering, Osaka University, 2-6 Yamadaoka, \\
Suita, Osaka 565-0871, Japan}

\author{Tsuyoshi \textsc{Inoue}\altaffilmark{1,9}}
\altaffiltext{9}{Department of Physics, Konan University, Okamoto 8-9-1, Kobe, Japan}

\author{Shuta J. \textsc{Tanaka}\altaffilmark{7,10}}
\altaffiltext{10}{Graduate School of Engineering, Osaka University, 2-1 Yamadaoka, Suita, Osaka, 565-0871, Japan}

%%% end:list of authors

%% `\KeyWords{}' always has to be placed before ``\maketitle'' 
%%  List of Key Words:  https://academic.oup.com/pasj/pages/Pasj_Keywords 

\maketitle

\begin{abstract}
A novel collisionless shock jump condition is suggested by modeling the entropy production at the shock transition region.
We also calculate downstream developments of the atomic ionization balance and the ion temperature relaxation in supernova remnants (SNRs).
The injection process and subsequent acceleration of cosmic-rays (CRs) in the SNR shocks are closely
related to the formation process of the collisionless shocks.
The formation of the shock is caused by wave-particle interactions.
Since the wave-particle interactions result in energy exchanges between electromagnetic fields and charged
particles, the randomization of particles associated with the shock transition may occur with the rate given by the scalar
product of the electric field and current.
We find that order-of-magnitude estimates of the randomization with reasonable strength of the electromagnetic
fields in the SNR constrain the amount of the CR nuclei and ion temperatures. The constrained amount of the CR nuclei can
be sufficient to explain the Galactic CRs. The ion temperature becomes significantly lower than in the case of no CRs.
To distinguish the case without CRs,
we perform synthetic observations of atomic line emissions
from the downstream region of the SNR RCW~86.
Future observations by {\it XRISM} and {\it Athena} can distinguish whether the SNR
shock accelerates the CRs or not from the ion temperatures.
\end{abstract}
%\linenumbers

\section{Introduction}
\label{sec:intro}
Collisionless shocks of supernova remnants (SNRs) are invoked as the primary sources
of Galactic cosmic-rays (CRs); however, the production process of CRs is an unsettled issue despite
numerous studies reported. The most generally accepted and widely studied
mechanism for CR acceleration is the diffusive shock acceleration~(DSA, \cite{bell78,blandford78}).
In the DSA mechanism, we assume energetic particles around the shock, and
the particles go into bouncing back and forth between the upstream and downstream
regions by scattering particles. The particle scattering results from interactions between plasma waves
and the particles. The maximum energy of the accelerated particles depends on
the magnetic field strength and turbulence~(e.g. \cite{lagage83a,lagage83b}).
To explain the energy spectrum of the CR nuclei observed around the Earth, the maximum energy of the accelerated protons
should be at least $10^{15.5}~{\rm eV}$ (so-called the knee energy). The knee energy can be achieved in the DSA mechanism
by the magnetic field strength of $\gtrsim100~{\rm \mu G}$ which is larger
than the typical strength of $\sim 1~{\rm \mu G}$ seen in the interstellar medium~(ISM, \cite{myers78,
beck01}).
\citet{bell04} pointed out that the upstream magnetic field is amplified by the effects of a back reaction from the
accelerated protons themselves. This amplification is called the Bell instability, whose
growth rate is proportional to the CR energy density.
Observations of nonthermal X-ray emissions around the SNR shocks imply the existence of amplified
magnetic fields in the downstream region~(e.g., \cite{vink03,bamba05,uchiyama07}).
Hence, in a modern scenario of the CR acceleration, the SNR shock is {\it assumed} to inject a considerably
large amount of CRs ($\gtrsim10$~\% of the shock kinetic energy), and their effects on the background plasma
are regarded as one of the most important issues. Since the wave-particle interactions also give
the formation of the collisionless shock, the injection of energetic particles, subsequent acceleration
by the DSA mechanism, and the amplification of the magnetic field are closely related to the formation
process.
Although many kinetic simulations studying collisionless shock physics are reported~(e.g., \cite{ohira13,
ohira16,ohira16b,matsumoto17,caprioli20,marcowith20}), self-consistent treatment of the collisionless shock, including
these effects, is currently incomplete due to the limitation of too short simulation time
compared to actual SNR shocks.
\par
When the SNR shock consumes its kinetic energy to accelerate energetic, nonthermal particles, the
downstream thermal energy can be lower than the case of adiabatic shock without
the CRs~(e.g., \cite{hughes00},
\cite{helder09,morlino13,morlino13b,morlino14,hovey15,hovey18,shimoda15,shimoda18a}).
Thus, we observe small downstream ion temperatures if the SNR shocks efficiently accelerate
the CR nuclei (protons and heavier ions).
In the near future, spatially resolved high energy-resolution spectroscopy of the SNR shock regions will be
achieved by the micro-calorimeter array with {\it Resolve} \citep{ishisaki18} onboard {\it XRISM} \citep{xrism20} and
with the {\it X-IFU} onboard {\it Athena} \citep{barret18} providing precise line diagnostics of plasmas to represent
the effect of the CR acceleration.
Note that observations of $\gamma$-ray emissions possibly provide the amount of the CR proton from luminosities.
However it may be challenging to determine the amounts of the CR nuclei individually.
Thus, in this paper, we study the shock jump conditions of ions, including the effects of CR acceleration.
To distinguish the case without the CRs by the future X-ray spectroscopy, we calculate temporal evolutions of
the downstream ionization structure and downstream ion temperatures resulting from the Coulomb interactions.
From the calculations of the downstream values, we also perform synthetic
observations of atomic lines, including effects of downstream turbulence.
Since the turbulence affects the line width by the Doppler effect, it is non-trivial whether
the observed line width reflects the intrinsic ion temperature.
\par
Typical example of the missing thermal energy measurement was provided at the SNR~RCW~86 (\cite{helder09}).
The RCW~86 is considered as the remnant of SN~185~\citep{vink06}, so its age is $\sim2000$~yr.
The current radius is $\sim15$~pc, and the shock velocity is $\sim3000~{\rm km~s^{-1}}$ with
an assumed distance of $2.5$~kpc~\citep{yamaguchi16}, where we estimate the angular size as $\sim40$~arcmin.
Since $2000~{\rm yr}\times3000~{\rm km~s^{-1}}\simeq 6~{\rm pc}<15~{\rm pc}$,
the blast wave has already decelerated. The mean expansion speed should be $\sim10^9~{\rm cm~s^{-1}}$
if the radius becomes $\sim10$~pc within the time of $\sim1$~kyr.
Such high expansion speed can be maintained during $1$~kyr if the progenitor star explodes in a wind-brown
cavity created by the progenitor system. \citet{broersen14} studied this
scenario by comparing the X-ray observations and hydrodynamical simulations. They concluded that the progenitor
star of SN~185 exploded as a Type-Ia supernova inside the wind-cavity. The progenitor system may be a binary system
consisting of a white-dwarf and donor star. In the present day, the RCW~86 shows H$\alpha$ filaments
everywhere~\citep{helder13}. The H$\alpha$ emission means that the shock is now propagating into a partially
ionized medium~\citep{chevalier80}. In the case of a stellar wind by a massive star, the ionization
front precedes the front of the swept-up matter~(e.g., \cite{arthur11}). Thus, the forward shock of the RCW~86 currently
may propagate in the medium not swept up by the wind. In this paper, we suppose such a scenario for the RCW~86
and perform synthetic observations of the X-ray atomic lines.
\par
This paper is organized as follows: In section~\ref{sec:model}, we review a physical model
of the temporal evolutions of the downstream ionization balance and the ion internal energies. The
ion temperatures are derived from the equation of state. Section~\ref{sec:shock} provides
shock jump conditions as initial conditions for the downstream temporal evolution. We introduce
shock jump conditions usually supposed in the SNRs and a novel condition given by modeling the entropy productions
of the ions due to the wave-particle interactions. The latter includes the effects of the CR acceleration, magnetic-field
amplification, and ion heating balance. The results of the downstream temporal evolutions
are summarized in section~\ref{sec:results}. In section~\ref{sec:synthetic}, we perform synthetic
observations of atomic lines, including the effects of the downstream turbulence. Finally, we summarize
our results and prospects.

\section{Physical Model of downstream ionization balance and ion internal energies}
\label{sec:model}
Here we review a physical model of the temporal evolutions of the downstream ionization
balance and ion internal energy. Let $V$ be a fluid parcel volume. The parcel contains
a mass of $M$. We assume that species within the parcel always have the Maxwell velocity
distribution function with a temperature of $T_j$, where the subscript $j$
indicates the species $j$. Then, the internal energy $E_j$ and pressure $P_j$
of the species $j$ can be written as
%
%%%%%%%%%%%%%%%%
\begin{eqnarray}
&& E_j = \frac{N_j k T_j}{\gamma-1}, \\
&& P_j = n_j k T_j,
\end{eqnarray}
%%%%%%%%%%%%%%%%
%
where $N_j$, $n_j$, and $k$ are the total number of the species $j$,
the number density of the species $j$ ($n_j=N_j/V$), and Boltzmann constant,
respectively. The adiabatic index is $\gamma=5/3$. From the first law of thermodynamics, we obtain
%
%%%%%%%%%%%%%%%%
\begin{eqnarray}
%&& \frac{d}{dt}\left(\frac{N_j kT_j}{\gamma-1} \right)
&& \frac{d E_j}{dt}
+ P_j \frac{dV}{dt}
= \frac{ d Q_j }{ dt },
\label{eq:1st law}
\end{eqnarray}
%%%%%%%%%%%%%%%%
%
where $dQ_j/dt$ is the external energy gain or loss per unit time of the species $j$;
we discuss it later.
Defining the internal energy per unit volume
$\varepsilon_j\equiv E_j/V$ and the external energy gain or loss per unit time per unit volume
$\dot{q}_j\equiv V^{-1}dQ_j/dt$, we rewrite the equation~(\ref{eq:1st law}) as
%
%%%%%%%%%%%%%%%%%
\begin{eqnarray}
\frac{d\varepsilon_j}{dt} = \dot{q_j} + \frac{\gamma\varepsilon_j}{\rho}\frac{d\rho}{dt},
\label{eq:de/dt}
\end{eqnarray}
%%%%%%%%%%%%%%%%%
%
where $\rho\equiv M/V$ is the total mass density.
In this paper, we suppose the case of young
SNRs and approximate their dynamics by the Sedov-Taylor model~\citep{sedov59,vink12}.
Then, we approximate the downstream velocity profile as
%
%%%%%%%%%%%%%%%%%
\begin{eqnarray}
v(r,t) = \left(1-\frac{1}{r_c}\right)\frac{V_{\rm sh}(t)}{R_{\rm sh}(t)}r,
\end{eqnarray}
%%%%%%%%%%%%%%%%%
%
where $r$ is the radial distance from the explosion center and $r_c$ is the compression ratio,
respectively. The radius of the SNR and the shock velocity are given by, respectively,
%
%%%%%%%%%%%%%%%%%
\begin{eqnarray}
&& R_{\rm sh}(t) = R_0 \left( \frac{ t }{ t_0 } \right)^{2/5}, \\
&& V_{\rm sh}(t) = \frac{dR_{\rm sh}}{dt} = \frac{2}{5}\frac{R_0}{t_0}\left(\frac{t}{t_0}\right)^{-3/5},
\end{eqnarray}
%%%%%%%%%%%%%%%%%
%
%
where we have assumed the ambient density structure around the SNR is uniform. The dimensional constants
$R_0$ and $t_0$ are characterized by the combination of the explosion energy of the supernova, the structure of
the ejecta, and the ambient density structure. The actual values of $R_0$ and $t_0$
are not used in our model calculation; we only use $V_{\rm sh}/R_{\rm sh}=(2/5)t^{-1}$.
The temporal evolution of the mass density along the trajectory of the fluid parcel is derived from
the continuous equation as
%
%%%%%%%%%%%%%%%%%%
\begin{eqnarray}
\frac{d\rho}{dt} = -\rho\frac{1}{r^2}\frac{\partial}{\partial r}\left(r^2 v\right)
= - \frac{6}{5}\left(1-\frac{1}{r_c}\right)\frac{ \rho }{ t }.
\label{eq:continuus}
\end{eqnarray}
%%%%%%%%%%%%%%%%%%
%
To calculate $\rho$ and $\varepsilon_j$ along the trajectory of the fluid parcel,
we introduce the position of the fluid parcel at $\tilde{r}(t)$ that is derived from the differential
equation of
%
%
%%%%%%%%%%%%%%%%%
\begin{eqnarray}
\frac{d\tilde{r}}{dt} = v(\tilde{r}(t),t) = \frac{2}{5}\left(1-\frac{1}{r_c}\right)\frac{\tilde{r}}{t}.
\end{eqnarray}
%%%%%%%%%%%%%%%%%
%
Defining the time $t_*$ when the fluid parcel currently at $\tilde{r}(t)=r$ crosses the shock,
i.e., $\tilde{r}(t_*)=R_{\rm sh}(t_*)$, we obtain
%
%%%%%%%%%%%%%%%%%
\begin{eqnarray}
\ln\frac{\tilde{r}(t)}{R_{\rm sh}(t_*)}
=\frac{2}{5}\left(1-\frac{1}{r_c}\right)\ln\frac{t}{t_*},
\end{eqnarray}
%%%%%%%%%%%%%%%%%
%
where we regard $r_c=$~const. When we observe the atomic line emissions from the fluid parcel at
$r=\tilde{r}(t_{\rm age})$, where $t_{\rm age}$ is the age of the SNR,
the crossing time is derived as \footnote{The compression ratio is strictly a function of the shock velocity.
Only for the calculation of $t_*$, we use the compression ratio given by $V_{\rm sh}(t_{\rm age})$, but for
the other cases, we calculate the shock jump conditions using $V_{\rm sh}(t_*)$ given by the calculated $t_*$.}
%
%%%%%%%%%%%%%%%%%
\begin{eqnarray}
t_*(r) = \left[ \frac{r}{ R_{\rm sh}{ (t_{\rm age}) }} \right]^{r_c} t_{\rm age}.
\label{eq:t_*}
\end{eqnarray}
%%%%%%%%%%%%%%%%%
%
%
Thus, by introducing $t'=t-t_*$, the temporal evolution of the downstream internal
energy and the mass density are written as, respectively,
%
%%%%%%%%%%%%%%%%%
\begin{eqnarray}
&& \frac{d\varepsilon_j}{dt'}
=\dot{q}_j - \frac{6\gamma}{5}\left(1-\frac{1}{r_c}\right)\frac{\varepsilon_j}{t'+t*},
\\
&& \rho(t') = \rho(t_*)\left(1+\frac{t'}{t_*}\right)^{-\frac{6}{5}\left(1-\frac{1}{r_c}\right)}.
\end{eqnarray}
%%%%%%%%%%%%%%%%%
%
Integrating the differential equation of $\varepsilon_j$ from $t'=0$ to $t'=t_{\rm age}-t_*(r)$ with the shock jump
conditions given by $V_{\rm sh}(t_*)$, we obtain the spatial profile of the downstream internal energy
at the observed time $t=t_{\rm age}$. The age is known for a historical SNR (e.g., SNR RCW~86, SN~1006, {\it Tycho}'s SNR,
{\it Kepler}'s SNR). The shock velocity at the current time can be estimated from the proper motion
of the shock. To calculate the rate of the Coulomb interactions (see below), the number density needs.
The density is evaluated from the surface brightness of the X-ray or H$\alpha$ emissions, for instance.
\par
Here we consider the energy source or sink term $\dot{q}_j$.
The charged particles exchange their momenta and energies via the Coulomb collision.
Although the exchange is negligible during the shock transition, the effect becomes important for the
long-time evolution in the downstream region. The energy exchange rate is given by~(e.g., \cite{spitzer62,itoh84})
%
%
%%%%%%%%%%%%%%%%%%
\begin{eqnarray}
\dot{q}_{j,{\rm Col}}
=
\sum_{m}
\frac{ \left( n_j\varepsilon_m - n_m\varepsilon_j \right)z_m{}^2 z_j{}^2\ln\Lambda }{5.87A_m A_j}
\left[\frac{T_m}{A_m}+\frac{T_j}{A_j} \right]^{-3/2},
\end{eqnarray}
%%%%%%%%%%%%%%%%%%
%
where $A_j$ and $\ln\Lambda$ are the particle mass in atomic mass units and
the Coulomb logarithm. In this paper, we fix $\ln\Lambda=30$ for simplicity.
For atoms, the energy transfer due to the ionization or recombination may be
given by
%
%
%%%%%%%%%%%%%%%%%%
\begin{eqnarray}
\dot{q}_{Z,z}
&=&
n_{\rm e}\left[
  R_{Z,z-1}\varepsilon_{Z,z-1}
-  \left( R_{Z,z}+K_{Z,z} \right)\varepsilon_{Z,z}
\right. \nonumber \\
&+&\left.
+K_{Z,z+1}\varepsilon_{Z,z+1}
\right].
\end{eqnarray}
%%%%%%%%%%%%%%%%%
%
where we introduce the notation $j=\{Z,z\}$ to represent the species with an atomic number $Z$ and
ionic charge state $z$, respectively (e.g., $Z=2$ and $z=1$ indicate He$^{+1}$ or He\emissiontype{II}). The
subscript `e' indicates the electron. The electron-impact ionization rate per unit
time per particle (${\rm s^{-1}~cm^3}$) is $R_{Z,z}(T_{\rm e})$, and the recombination rate per unit time per particle
is $K_{Z,z}(T_{\rm e})$. In this paper, we omit the charge-exchange reactions and the ion impact ionization
for simplicity and consider ten atoms H, He, C, N, O, Ne, Mg, Si, S, and Fe with the solar abundance~\citep{asplund09}.
The atomic data used in this paper are the same as \citet{shimoda21}:
The ionization cross-sections are given by \citet{janev93}
for H, and \citet{lennon88} for the others. The fitting functions for those data
are given by International Atomic Energy Agency.\footnote{$\langle$https://www.iaea.org/resources/databases/aladdin$\rangle$}
Table~\ref{tab:recombination} summarises the literature on the recombination rates. Those data are fitted by the
Chebyshev polynomials with twenty terms. For the hydrogen-like atoms, the fitting function is given by \citet{kotelnikov19}.
The electron number density is given by the charge neutrality condition as
%
%%%%%%%%%%%%%%%%%
\begin{eqnarray}
n_{\rm e} = \sum_{Z}\sum_{z=1}^{z=Z} z n_{Z,z},
\label{eq:ne}
\end{eqnarray}
%%%%%%%%%%%%%%%%%
%
and the total number density $n$ is given by
%
%%%%%%%%%%%%%%%%%
\begin{eqnarray}
n = n_{\rm e} + \sum_{Z} \sum_{z=0}^{z=Z} n_{Z,z}.
\label{eq:n}
\end{eqnarray}
%%%%%%%%%%%%%%%%%
%
For electrons, by following \citet{shimoda21}, the radiative and ionization losses are given by
%
%%%%%%%%%%%%%%%%%
\begin{eqnarray}
\dot{q}_{\rm e}
= - \sum_Z\sum_{z=0}^{Z-1} n_{\rm e}n_{Z,z}R_{Z,z}I_{Z,z}
  - n_{\rm e}\sum_{Z,z}n_{Z,z} W_{Z,z},
\end{eqnarray}
%%%%%%%%%%%%%%%%%
%
where $I_{Z,z}$ is the first ionization potential of the species $j=\{Z,z\}$
(we omit the inner shell ionization). The radiation power $W_{Z,z}$ includes
the bound-bound, free-bound, free-free, and two-photon decay.
For the continuum components, the formula given by \citet{gronen78} (free-free and two-photon decays) and
\citet{mewe86} (free-bound) are used. For the bound-bound component, the radiation power per particle
is given by
%
%%%%%%%%%%%%%%%%%
\begin{eqnarray}
W_{Z,z}^{ul} = E_{ul}C_{lu},
\end{eqnarray}
%%%%%%%%%%%%%%%%%
%
where the emitted photon energy is the subtraction of the upper energy level $E_u$
and the lower energy level $E_l$, $E_{ul}=E_u-E_l$. The collisional excitation rate
per unit time per particle (${\rm s^{-1}~cm^3}$) is given by~(e.g., \cite{Osterbrock06})
%
%%%%%%%%%%%%%%
\begin{eqnarray}
&& C_{lu} = 8.629\times10^{-6}
          \frac{ \Omega_{lu} }{ g_l }
          \frac{ {\rm e}^{-\frac{ E_{ul} }{ kT_{\rm e} } } }{ \sqrt{ T_{\rm e} } },
\end{eqnarray}
%%%%%%%%%%%%%%
%
where $g_l$ is the statistical weight of the lower level.
The collision strength is 
%
%%%%%%%%%%%%%%
\begin{eqnarray}
&& \Omega_{lu}
          = \frac{ 8\pi }{ \sqrt{3} }
            \frac{ g_l f_{lu} }{ E_{ul,{\rm Ryd}} } 
            \bar{g}(T_{\rm e}),
\end{eqnarray}
%%%%%%%%%%%%%%
%
where $f_{lu}$ is the oscillator strength, and $E_{ul,{\rm Ryd}}$ is the photon
energy given in the Rydberg unit. The $g_l$ is the statistical weight of 
the lower level. The averaged Gaunt factor is $\bar{g}$. The value of the averaged
Gaunt factor is around unity and determines the detailed temperature dependence
of the excitation rate. The precise data of the excitation rate (or $\bar{g}$)
are, however, not available yet. The following fitting function
\citep{mewe72}
%
%%%%%%%%%%%%%%
\begin{eqnarray}
\bar{g}(T_{\rm e}) = 0.15
+ 0.28
\left[ \log \left( \frac{ \chi+1 }{\chi} \right)
     - \frac{ 0.4 }{ (1+\chi)^2 }
\right],
\end{eqnarray}
%%%%%%%%%%%%%%
%
where $\chi = E_{ul}/kT_{\rm e}$, is used for the neutral atoms, while $\bar{g}=1$ is assumed for the
ionized atoms. Note that the cooling function mainly depends on the ionization structure
rather than $\bar{g}$. For the oscillator strength and energy levels,
the data table given by the National Institute of Standards and
Technology\footnote{$\langle$https://www.nist.gov/pml/atomic-spectra-databasea$\rangle$} are used.
For the calculation of the radiative cooling rate, it is sufficient to consider only
the allowed transitions from the ground state.
We obtain the net radiation power and thus the net radiative loss by integrating the photon frequency.
Here is a summary of the energy source or sink term:
$\dot{q}_j=\dot{q}_{Z,0}$ for the neutral atoms,
$\dot{q}_j=\dot{q}_{j,{\rm Col}}+\dot{q}_{Z,z}$ for the ions, and
$\dot{q}_j=\dot{q}_{j,{\rm Col}}+\dot{q}_{\rm e}$ for the electrons.

%
%%%%%%%%%%%%%%
\begin{table*}[htbp]
\tbl{Literature on the recombition rate. The superscript $^*$
denotes that we use the Mewe's formula for the radiative recombination
\citep{mewe80a,mewe80b}.}{%
\tabcolsep 3pt
\begin{tabular}{cc cc cc}
\hline
Ion & Literature & Ion & Literature & Ion & Literature \\
\hline
 C$^{+1 }$ & \citet{nahar99}          &  Mg$^{+5 }$ & \citet{arnaud85}         &   S$^{+12}$ & \citet{mewe80a,mewe80b}  \\
 C$^{+2 }$ & \citet{nahar99}          &  Mg$^{+6 }$ & \citet{zatsarinny04}     &   S$^{+13}$ & \citet{mewe80a,mewe80b}  \\
 C$^{+3 }$ & \citet{nahar97}          &  Mg$^{+7 }$ & \citet{nahar95}          &   S$^{+14}$ & \citet{arnaud85}         \\
 C$^{+4 }$ & \citet{nahar97}          &  Mg$^{+8 }$ & \citet{arnaud85}         &   S$^{+15}$ & \citet{arnaud85}         \\
 C$^{+5 }$ & \citet{nahar97}          &  Mg$^{+9 }$ & \citet{arnaud85}         &  Fe$^{+1 }$ & \citet{nahar97}          \\
 N$^{+1 }$ & \citet{zatsarinny04}     &  Mg$^{+10}$ & \citet{arnaud85}         &  Fe$^{+2 }$ & \citet{nahar97}          \\
 N$^{+2 }$ & \citet{nahar97}          &  Mg$^{+11}$ & \citet{arnaud85}         &  Fe$^{+3 }$ & \citet{nahar97}          \\
 N$^{+3 }$ & \citet{nahar97}          &  Si$^{+1 }$ & \citet{nahar00}          &  Fe$^{+4 }$ & \citet{nahar98}          \\
 N$^{+4 }$ & \citet{nahar97}          &  Si$^{+2 }$ & \citet{altun07}          &  Fe$^{+5 }$ & \citet{nahar99}          \\
 N$^{+5 }$ & \citet{nahar06}          &  Si$^{+3 }$ & \citet{mewe80a,mewe80b}  &  Fe$^{+6 }$ & \citet{arnaud85}         \\
 N$^{+6 }$ & \citet{nahar06}          &  Si$^{+4 }$ & \citet{zatsarinny03}     &  Fe$^{+7 }$ & \citet{nahar00}          \\
 O$^{+1 }$ & \citet{nahar98}          &  Si$^{+5 }$ & \citet{zatsarinny06}$^*$ &  Fe$^{+8 }$ & \citet{arnaud85}         \\
 O$^{+2 }$ & \citet{zatsarinny04}     &  Si$^{+6 }$ & \citet{zatsarinny03}     &  Fe$^{+9 }$ & \citet{arnaud85}         \\
 O$^{+3 }$ & \citet{nahar98}          &  Si$^{+7 }$ & \citet{mitnik04}$^*$     &  Fe$^{+10}$ & \citet{lestinsky09}$^*$  \\
 O$^{+4 }$ & \citet{nahar98}          &  Si$^{+8 }$ & \citet{zatsarinny04}     &  Fe$^{+11}$ & \citet{novotny12}$^*$    \\
 O$^{+5 }$ & \citet{nahar98}          &  Si$^{+9 }$ & \citet{nahar95}          &  Fe$^{+12}$ & \citet{hahn14}$^*$       \\
 O$^{+6 }$ & \citet{nahar98}          &  Si$^{+10}$ & \citet{arnaud85}         &  Fe$^{+13}$ & \citet{arnaud85}         \\
 O$^{+7 }$ & \citet{nahar98}          &  Si$^{+11}$ & \citet{arnaud85}         &  Fe$^{+14}$ & \citet{altun07}$^*$      \\
Ne$^{+1 }$ & \citet{arnaud85}         &  Si$^{+12}$ & \citet{arnaud85}         &  Fe$^{+15}$ & \citet{murakami06}$^*$   \\
Ne$^{+2 }$ & \citet{zatsarinny03}     &  Si$^{+13}$ & \citet{arnaud85}         &  Fe$^{+16}$ & \citet{zatsarinny04}     \\
Ne$^{+3 }$ & \citet{mitnik04}$^*$     &   S$^{+1 }$ & \citet{mewe80a,mewe80b}  &  Fe$^{+17}$ & \citet{arnaud85}         \\
Ne$^{+4 }$ & \citet{zatsarinny04}     &   S$^{+2 }$ & \citet{nahar95}          &  Fe$^{+18}$ & \citet{zatsarinny03}     \\
Ne$^{+5 }$ & \citet{nahar95}          &   S$^{+3 }$ & \citet{nahar00}          &  Fe$^{+19}$ & \citet{savin02}$^*$      \\
Ne$^{+6 }$ & \citet{arnaud85}         &   S$^{+4 }$ & \citet{altun07}          &  Fe$^{+20}$ & \citet{zatsarinny04}     \\
Ne$^{+7 }$ & \citet{arnaud85}         &   S$^{+5 }$ & \citet{arnaud85}         &  Fe$^{+21}$ & \citet{arnaud85}         \\
Ne$^{+8 }$ & \citet{nahar06}          &   S$^{+6 }$ & \citet{zatsarinny04}     &  Fe$^{+22}$ & \citet{arnaud85}         \\
Ne$^{+9 }$ & \citet{nahar06}          &   S$^{+7 }$ & \citet{zatsarinny06}$^*$ &  Fe$^{+23}$ & \citet{mewe80a,mewe80b}  \\
Mg$^{+1 }$ & \citet{mewe80a,mewe80b}  &   S$^{+8 }$ & \citet{zatsarinny03}     &  Fe$^{+24}$ & \citet{nahar01}          \\
Mg$^{+2 }$ & \citet{zatsarinny04}     &   S$^{+9 }$ & \citet{mitnik04}$^*$     &  Fe$^{+25}$ & \citet{nahar01}          \\
Mg$^{+3 }$ & \citet{arnaud85}         &   S$^{+10}$ & \citet{zatsarinny04}     &             &                          \\
Mg$^{+4 }$ & \citet{zatsarinny04}     &   S$^{+11}$ & \citet{nahar95}          &             &                          \\
\hline
\end{tabular} }
\label{tab:recombination}
\end{table*}
%%%%%%%%%%%%%%
%
\par
Since the SNR shock may heat the plasma faster than the Coulomb collisions due to the wave-particle
interactions in the plasma, the ionization state of atoms can significantly deviate from the ionization
equilibrium. Thus, we simultaneously solve the atomic rate equations
%
%%%%%%%%%%%%%%%%%
\begin{eqnarray}
\frac{d}{ dt' }\left( \frac{ n_{Z,z} }{ \rho } \right)
&=& n_{\rm e}
\left[   R_{Z,z-1} \frac{ n_{Z,z-1} }{\rho}
-   \left( R_{Z,z} + K_{Z,z} \right) \frac{ n_{Z,z} }{\rho}
\right. \nonumber \\
&+& \left.
K_{Z,z+1} \frac{ n_{Z,z+1} }{\rho}
\right].
\label{eq:rate eq}
\end{eqnarray}
%%%%%%%%%%%%%%%%%
%
Note that in our formulation, the velocity distribution function of the species is always
assumed to be the Maxwellian.

\section{Shock jump conditions}
\label{sec:shock}
Here we give the initial conditions for the temporal developments of the downstream
ionization balance and temperature relaxation by considering shock jump conditions.
We introduce the conditions usually supposed in the SNR shocks from analogs of
collisional shocks and the novel condition given by modeling the energy exchange between
electromagnetic fields and particles.

\subsection{collisional shock model (Model~0, 1, and 2)}
For the pre-shock gas (denoted by the subscript `0'), we set $T_{j,0}=T_0=3\times10^4$~K with
assuming the collisional ionization equilibrium and temperature equilibrium. In this condition, the fraction
of the neutral atoms is $\sim1.3\times10^{-2}$ in the number.
For the downstream values (denoted by the subscript `2') of the charged particles, assuming a negligibly
small magnetic field at the upstream region (or a parallel shock), we consider the total flux conservation
laws as
%
%%%%%%%%%%%%%%
\begin{eqnarray}
\rho_{0} v_{0} &=& \rho_{2} v_{2}, \\
\rho_{0}v_0{}^2+P_{0}
&=& \rho_{2}v_{2}{}^2+P_{2}, \\
\frac{ \rho_0 v_0{}^3 }{2} + \left(\varepsilon_{0}+P_{0}\right)v_{0}
&=&
\frac{ \rho_2 v_{2}{}^3 }{2} + \left(\varepsilon_{2}+P_{2}\right)v_2,
\end{eqnarray}
%%%%%%%%%%%%%%
%
where the total pressure is  $P=\sum_j P_j$, and the total internal energy is $\varepsilon=\sum_j\varepsilon_j$, respectively.
The mass density of the species $j$ is $\rho_j=m_j n_j$, where $m_j$ is the particle mass, and the total mass density is $\rho=\sum_j\rho_j$.
The compression ratio $r_c$ and total pressure jump $x_c$ are derived as
%
%%%%%%%%%%%%%%
\begin{eqnarray}
&& r_c \equiv \frac{\rho_2}{\rho_0} = \frac{v_0}{v_2}
= \frac{ \left(\gamma+1\right){\cal M}_{\rm s}{}^2 }{ \left(\gamma-1\right){\cal M}_{\rm s}{}^2+2 }
\label{eq:rc ncr} \\
&& x_c \equiv \frac{ P_2 }{ P_0 } = \frac{ \varepsilon_2}{ \varepsilon_0 } 
= \frac{ 2\gamma{\cal M}_{\rm s}{}^2 - \left(\gamma-1\right) }{ \gamma+1 },
\label{eq:xc ncr}
\end{eqnarray}
%%%%%%%%%%%%%%
%
where ${\cal M}_{\rm s}\equiv v_0/\sqrt{\gamma P_0/\rho_0}$ is the sonic Mach number defined by
the total pressure and mass density. For each species $j$, we assume the flux conservation laws as
%
%%%%%%%%%%%%%%
\begin{eqnarray}
\rho_{j,0} v_{0} &=& \rho_{j,2} v_{2}, \\
\rho_{j,0}v_0{}^2+P_{j,0}
&=& \rho_{j,2}v_{2}{}^2+P_{2}\frac{\rho_{j,0}}{\rho_0}, \\
\frac{ \rho_{j,0}v_0{}^3 }{2} + \left(\varepsilon_{j,0}+P_{j,0}\right)v_0
&=&
\frac{ \rho_{j,2}v_{2}{}^3 }{2}
+ \left\{\left(\varepsilon_{j,2}+P_{j,2}\right)\frac{\rho_{j,0}}{\rho_0}\right\}v_2,~~~~~~
\end{eqnarray}
%%%%%%%%%%%%%%
%
where we have assumed that the downstream ion velocities are the same as each other ($v_{j,2}=v_2$), and
that the downstream internal energy $\varepsilon_{j,2}=(\rho_{j,0}/\rho_0)\varepsilon_2$ and pressure $P_{j,2}=(\rho_{j,0}/\rho_0)P_2$
are proportional to the upstream kinetic energy $\rho_{j,0}v_0{}^2/2$. The downstream temperature of the species $j$, $kT_{j,2}=
P_{j,2}/n_{j,2}=(P_2/\rho_2)m_j$, is derived as
%
%%%%%%%%%%%%%%
\begin{eqnarray}
kT_{j,2}
&=&\frac{m_j v_0{}^2}{r_c}\left(1-\frac{1}{r_c}+\frac{1}{\gamma{\cal M}_{\rm s}{}^2}\right) \nonumber \\
&=&\frac{ \left(\gamma-1\right) m_j v_0'{}^2 }{2}
\frac{ \left( {\cal M}_{\rm s}{}^2+\frac{2       }{\gamma-1} \right)
       \left( {\cal M}_{\rm s}{}^2-\frac{\gamma-1}{2\gamma } \right) }
     { \left( {\cal M}_{\rm s}{}^2 - 1 \right)^2 },~~~~~~
\label{eq:kT2 ncr}
\end{eqnarray}
%%%%%%%%%%%%%%
%
where $v_0'=v_0-v_2$ is the upstream velocity measured in the downstream rest. In the strong shock limit with $\gamma=5/3$, we obtain
the relation of $(3/2)kT_{j,2}=m_jv_0'{}^2/2$ indicating that the upstream coherent motion of the particles is completely
randomized due to the shock transition. The temperature ratio of the species $j$ to $k$
is equal to their ion mass ratio, $T_j/T_k = m_j/m_k$. This corresponds that the widths of the Maxwell velocity distribution
function of each species are the same.
The neutral particles do not form the shock structure because they do not interact with
the electromagnetic fields. Then, for the neutral particles, we approximately adopt
$\varepsilon_{j,2}=\rho_{j,0}v_0{}^2/2+\varepsilon_{j,0}$.
We will refer this collisional shock model to Model~0.
\par
To investigate the effects of the electron heating around the shock transition region, we parameterize
the energy exchange between protons and electrons as
%
%%%%%%%%%%%%%%
\begin{eqnarray}
&& \varepsilon_{{\rm p},2} = \varepsilon_2 \frac{ \rho_{{\rm p},0} }{ \rho_0 } \left(1-f_{\rm eq}\right), \\
&& \varepsilon_{{\rm e},2} = \varepsilon_2
\left( \frac{ \rho_{{\rm e},0} }{ \rho_0} + \frac{ \rho_{{\rm p},0} }{ \rho_0}f_{\rm eq} \right)
\end{eqnarray}
%%%%%%%%%%%%%%
%
where the subscript `p' denotes the proton. The degree of equilibrium is represented by the parameter
$f_{\rm eq}$ that is related to the temperature ratio as follows
%
%%%%%%%%%%%%%%
\begin{eqnarray}
\frac{ \varepsilon_{\rm e,2} }{ \varepsilon_{\rm p,2} }
=\frac{ n_{\rm e,2}  }{ n_{\rm p,2}  }
 \frac{ T_{\rm e,2}  }{ T_{\rm p,2}  }
=\frac{ (\rho_{\rm e,0}/\rho_{\rm p,0}) + f_{\rm eq} }{ 1-f_{\rm eq} }.
\end{eqnarray}
%%%%%%%%%%%%%%
%
Introducing $\beta\equiv T_{\rm e,2}/T_{\rm p,2}$, we obtain
%
%%%%%%%%%%%%%
\begin{eqnarray}
f_{\rm eq}  = \frac{ n_{\rm e,0} }{ n_{\rm p,0} }
\frac{ \beta - ( m_{\rm e}/m_{\rm p} ) }{ 1 + ( n_{\rm e,0}/n_{\rm p,0} )\beta }
\end{eqnarray}
%%%%%%%%%%%%%
%
We consider the cases of $\beta=0.01$ (Model~1) and $\beta=0.1$ (Model~2).
Although the electron might exchange its internal energy with other ions, we omit this possibility for simplicity.
Complete treatments of the electron heating around the shock may need to solve
the nature of electromagnetic fields and wave-particle interactions in detail, and this issue
is unsettled yet~(e.g., \cite{ohira07,ohira08,rakowski08,laming14}).

\subsection{collisionless shock model (Model~3, 4, and 5)}
Here we consider another way of giving a shock transition with the CR acceleration.
We assume that a part of shock kinetic energy is consumed for the generation of the CRs
and the amplification of the magnetic field. The generated magnetic field is assumed to be disturbed
(not an ordered field). In this model, we consider the randomization of the particles incoming from the
far upstream region at the shock transition region. The randomization is quantified by the entropy.
We notice that the `randomization' results in a more isotropic particle distribution downstream
than the pre-shock one (measured in the shock rest frame). In the collisional shocks, it may be called as the
`thermalization', however, the particle distribution may deviate from the Maxwellian in the collisionless shocks.
We use the `randomization' for both collisional and collisionless shocks in the following.
\par
Conservation laws of total mass and momentum flux can be written as
%
%%%%%%%%%%%%%
\begin{eqnarray}
\rho_0 v_0 &=& \rho_2 v_2, \\
\rho_0 v_0{}^2 + P_0 + F_{{\rm esc}}&=& \rho_2 v_2{}^2 + P_2 + \frac{\delta B^2}{4\pi} + P_{\rm cr},
\end{eqnarray}
%%%%%%%%%%%%%
%
where the generated (turbulent) magnetic-field strength is $\delta B$. We regard that the field with $\delta B$ has a
coherent length scale (injection scale of turbulence) much larger than the Larmor radius of the thermal particles with
a velocity of $\sim v_0$ and that the turbulence cascades to the smaller scale.
The disturbances associated with the field are assumed to randomize the thermal particles by the wave-particle interactions.
The CR pressure of the species $j$ is defined as  $P_{{\rm cr},j}$ and the total CR pressure is
$P_{\rm cr}=\sum_j P_{{\rm cr},j}$. The net momentum flux of escaping CRs is $F_{{\rm esc}} \lesssim \rho_0v_0{}^3/3c$.
We neglect the total flux and the flux of each species $j$ in this article ($F_{\rm esc}=0$ and $F_{{\rm esc},j}=0$).
For each species denoted by the subscript $j$, we give the flux conservation laws as
%
%%%%%%%%%%%%%
\begin{eqnarray}
\rho_{j,0} v_0 &=& \rho_{j,2} v_2, \\
\rho_{j,0} v_0{}^2 + P_{j,0}
&=& \rho_{j,2} v_2{}^2
+ P_{j,2}
\nonumber \\
&+&
\left(
\frac{\delta B^2}{4\pi}+P_{\rm cr}
\right)\frac{\rho_{j,0}}{\rho_0}
\end{eqnarray}
%%%%%%%%%%%%%
%
where we assume a contribution of the species $j$ for the magnetic field amplification and
nonthermal pressure is proportional to the upstream kinetic energy $\rho_{j,0}v_0{}^2/2$. The compression
ratios of the species $j$ are equal ($v_{2,j}=v_2$).  From these
conservation laws, we can derive the relation between the compression ratio $r_{c,j} =\rho_{j,2}/\rho_{j,0}=\rho_2/
\rho_0\equiv r_c$ and the jump of internal energy $x_{c,j}\equiv\varepsilon_{j,2}/\varepsilon_{j,0}$ as
%
%%%%%%%%%%%%%%
\begin{eqnarray}
r_c     &=& \left[ 1 + \frac{1-x_{c,j}}{\gamma {\cal M}_{{\rm s},j}{}^2} - \xi_{\rm B} - \xi_{{\rm cr}} \right]^{-1}, 
\label{eq:r}
\end{eqnarray}
%%%%%%%%%%%%%%
%
or
%
%%%%%%%%%%%%%%
\begin{eqnarray}
x_{c,j} &=& 1 + \gamma {\cal M}_{{\rm s},j}{}^2\left( 1 - \frac{1}{r_c} - \xi_{\rm B} - \xi_{{\rm cr}} \right),
\label{eq:x}
\end{eqnarray}
%%%%%%%%%%%%%%
%
where $\xi_{\rm B}\equiv\delta B^2/(4\pi\rho_0 v_0{}^2)$, $\xi_{{\rm cr}}\equiv P_{{\rm cr}}/(\rho_{0}v_0{}^2)$, and
${\cal M}_{{\rm s},j}=v_0/\sqrt{\gamma P_{j,0}/\rho_{j,0}}$ is the sonic Mach number defined by the pressure and density of the species $j$.
Thus, once another relation between $r_c$ and $x_{c,j}$ is found, we can derive the shock jump condition with given $\xi_{\rm B}$
and $\xi_{{\rm cr}}$. As usual, the energy flux conservation is considered by modeling
the magnetic field amplification and the injection rate of nonthermal
particles. Since we focus on the downstream ion temperature, we consider the randomization process of thermal ions rather
than modeling the behavior of nonthermal particles.
Thus, we consider the entropy production of the thermal particles explicitly.
\par
The entropy of the species $j$ per unit mass is defined as
%
%%%%%%%%%%%%%
\begin{eqnarray}
ds_j = \frac{1}{M_j}\frac{ d\tilde{Q}_j }{ kT_j },
\label{eq:dsj}
\end{eqnarray}
%%%%%%%%%%%%%
%
where $d\tilde{Q}_j$ is the energy transferred from electromagnetic fields to the {\it internal energy} of the species $j$
due to the shock transition, and $M_j=N_jm_j$ is the total mass of the species $j$ within the fluid parcel.\footnote{
This definition corresponds to a reversible process. The collisionless shock is formed by the wave-particle
interaction, which might be a reversible process like a plasma echo, for example. Although it may be an unsettled issue,
we apply this definition of entropy in this article. Note that the entropy is defined as a non-dimensional value
differing from the usual dimensional definition in thermodynamics, $dS=dQ/T$.}
Note that $d\tilde{Q}_j=dE_j+P_jdV$ indicates only the increment of the internal energy rather than the total kinetic energy
of the thermal particles (a sum of the bulk motion and the random motion). The upstream total kinetic energy of the thermal particles
is divided into $\delta B$ and $P_{\rm cr}$.
Substituting $d\tilde{Q}_j=dE_j+P_jdV$ to the equation~(\ref{eq:dsj}), and using the relation of
$d\varepsilon_j=d(\rho_j e_j)=e_j d\rho_j + \rho_jde_j$, where $e_j\equiv E_j/M_j$, we can derive
the change of the internal energy per unit volume as
%
%%%%%%%%%%%%%
\begin{eqnarray}
\frac{d\varepsilon_j}{\varepsilon_j}
=\gamma\frac{d\rho_j}{\rho_j}
+(\gamma-1)m_j  ds_j.
\end{eqnarray}
%%%%%%%%%%%%%
%
Note that we have presumed that $N_j$ is constant during the shock transition.
Thus, we obtain the entropy jump before and after the shock transition,
$\Delta s_j=s_{j,2}-s_{j,0}$ as
%
%%%%%%%%%%%%%
\begin{eqnarray}
\left(\gamma-1\right)m_j\Delta s_j
&=&\ln\left( \frac{ \varepsilon_{j,2} }{ \varepsilon_{j,0} } \right)
-\gamma\ln\left( \frac{ \rho_{j,2} }{ \rho_{j,0} } \right)
\nonumber \\
&=&\ln x_{c,j} - \gamma\ln\ r_c.
\label{eq:eos}
\end{eqnarray}
%%%%%%%%%%%%%
%
Then, the jump conditions are derived by estimating $\Delta s_j$ independently from the equation~(\ref{eq:eos}).
Since the SNR shock is expected to be formed by the wave-particle interactions, the transferred energy in total $\Delta\tilde{Q}_j$
may be around $\sim\bm{J}_j\cdot\bm{E}\Delta t_j$, where $\bm{J}_j$ is the electric current of species $j$.
The electric field measured in the comoving frame of the ions is $\bm{E}$.
$\Delta t_j$ is a time taking the shock transition.
We estimate each value as $J_j\sim q_j N_j\langle\tilde{v}_j\rangle$, $E\equiv|\bm{E}|\sim (\langle\tilde{v}_j\rangle/c)\delta B$,
and $\Delta t_j\sim m_jc/q_j\delta B$, where
$q_j$ is the electric charge of the species $j$,
$\langle\tilde{v}_j\rangle=v_0+\sqrt{2kT_0/m_j}$ is the mean kinetic velocity of the species $j$,
and $c$ is the speed of light, respectively.
The transition time scale is assumed to be comparable with an inverse of the cyclotron frequency. In a hybrid simulation
solving the particle acceleration~(e.g., \cite{ohira16}), the shock jump seems to occur at a very small length
scale despite a significant amplification of turbulent magnetic fields at the `upstream' region
(it may correspond to a shock precursor region in our situation). We regard that the randomization of particles resulting in
the shock transition mainly occurs at such a very small length scale. Thus, we assume the entropy production due
to the shock transition as
%
%%%%%%%%%%%%%
\begin{eqnarray}
\Delta s_j
&=& \frac{1}{M_j} \frac{ J_jE\Delta t_j }{ kT_j }
\nonumber \\
&=& \frac{1}{M_j}
\frac{q_jN_j\langle\tilde{v}_j\rangle}{kT_{j,2}}
\frac{\langle\tilde{v}_j\rangle}{c}\delta B
\frac{m_jc}{q_j\delta B}
\nonumber \\
&=&
\frac{\langle\tilde{v}_j\rangle^2 }{ kT_0 }
\frac{ r_c }{ x_{c,j} },
\label{eq:delta s}
\end{eqnarray}
%%%%%%%%%%%%%
%
where we suppose $T_j\sim T_{j,2}$.
Substituting the equition~(\ref{eq:delta s}) to the equation~(\ref{eq:eos}), we obtain
the relation between $r_c$ and $x_{c,j}$ as
%
%%%%%%%%%%%%%
\begin{eqnarray}
f\equiv
\frac{ x_{c,j} }{ r_c }
\left[
\ln x_{c,j} -\gamma\ln r_c
\right]
-\gamma\left(\gamma-1\right)
 \left( \frac{ \langle\tilde{v}_j\rangle }{ v_0 } \right)^2
 {\cal M}_{{\rm s},j}{}^2 = 0.
\nonumber \\
\label{eq:fx}
\end{eqnarray}
%%%%%%%%%%%%%
%
We solve this equation setting $P_{\rm cr}$, $\delta B$ and ${\cal M}_{{\rm s},j}=v_0/\sqrt{\gamma P_{j,0}/\rho_{j,0}}$
with the equation~(\ref{eq:r}) to derive $x_{c,j}$ in the case of the proton by regarding
that the most abundant ions form the shock structure.
Then, the compression ratio  $r_{c}$ is derived from the equation~(\ref{eq:r}) by using the
derived $x_{c,{\rm p}}$. The downstream pressures of the other species $j$ are derived
from the equation~(\ref{eq:x}) by using the derived compression ratio $r_c$.
Note that if we supposed small $\delta B$ and $P_{\rm cr}$,
the resultant downstream values would be different from the case of the collisional shock (Model~0) reflecting
the different randomization process.
In this paper, we consider the most efficiently accelerating CR shock feasible. In such a situation, the CR
pressure is a practical function of $\delta B$ because of the energy budget of the shock. The upstream kinetic energy
is divided into the thermal energy, the magnetic field, and the CRs. The fraction of the thermal energy is given by the entropy production.
The fraction of the magnetic field is treated as a free parameter. Thus, the remaining energy is divided into the CRs.
\par
%
%%%%%%%%%%%%%%
\begin{figure}[htbp]
\begin{center}
\includegraphics[scale=0.65]{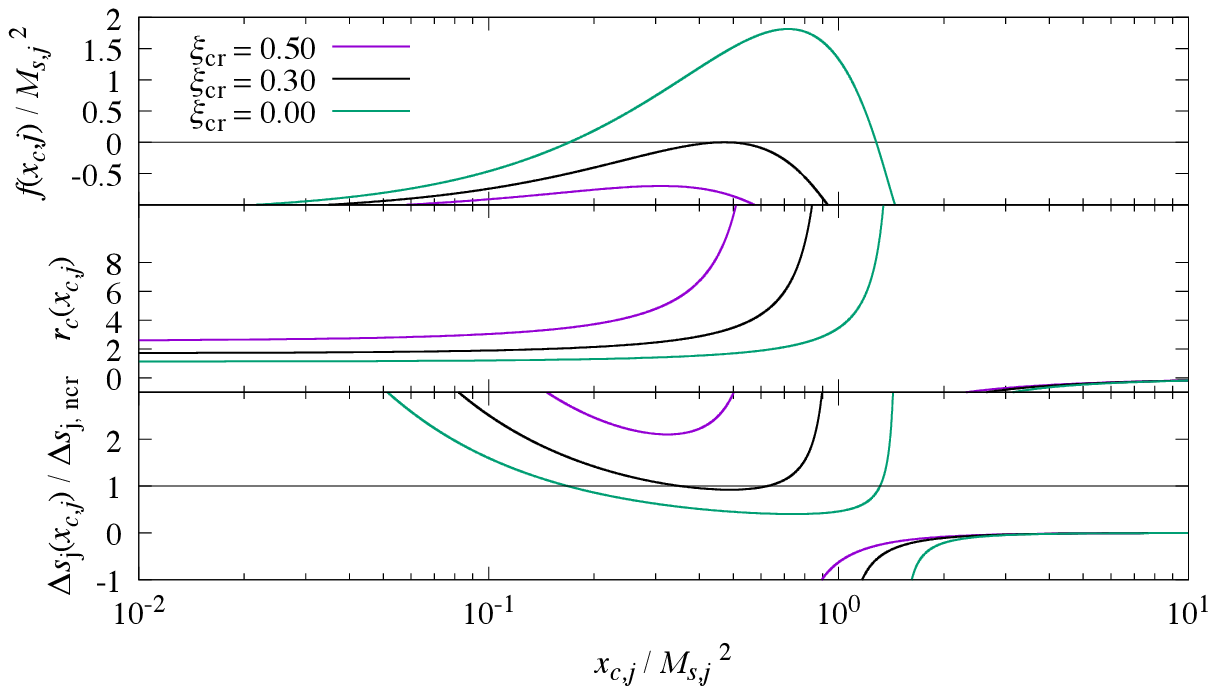}
\includegraphics[scale=0.65]{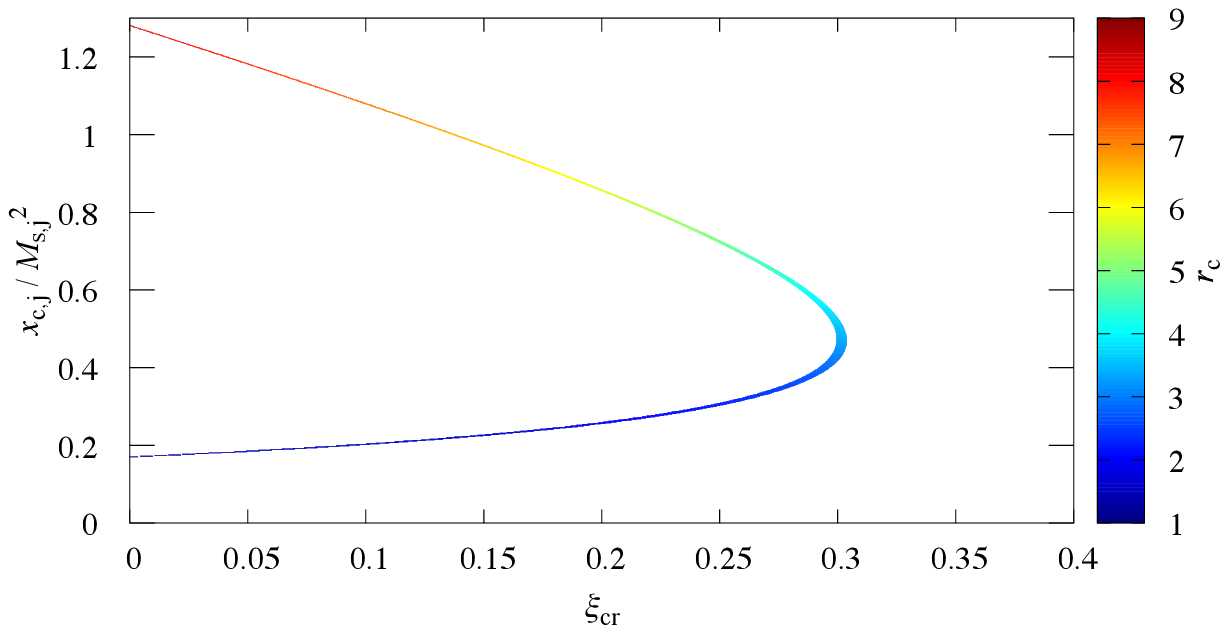}
\end{center}
\caption{
(left panel) The function of $f=f(x_{c,j})$ defined in the equation~(\ref{eq:fx}) (upper part), the compression
ratio $r_c=r_c(x_{c,j})$ (middle part), and $\Delta s_j(x_{c,}j)/\Delta s_{j,{\rm ncr}}$ (lower part) for the proton
with $\gamma=5/3$.
We set parameters as $\xi_{\rm cr}=0.5$ (purple line), $\xi_{\rm cr}=0.3$ (black line), and $\xi_{\rm cr}=0$ (green line)
with fixed values of $v_0=4000~{\rm km~s^{-1}}$, $T_0=3\times10^4$~K, and $1/\sqrt{\xi}_{\rm B}=v_0/(\delta B/\sqrt{4\pi\rho_0})=3$.
Note that ${\cal M}_{{\rm s},j}=197$.
(right panel) Solutions of $f=0$ with fixed $1/\sqrt{\xi}_{\rm B}=3$ and ${\cal M}_{{\rm s},j}=197$ for the proton.
The horizontal axis shows the CR fraction $\xi_{\rm cr}$ and the vertical axis shows the pressure jump $x_{{\rm c},j}/{\cal M}_{{\rm s},j}{}^2$.
The color indicates the compression ratio $r_c$.}
\label{fig:fx}
\end{figure}
%%%%%%%%%%%%%%
%
The left panel of figure~\ref{fig:fx} shows $f=f(x_{c,j})$ (upper part), $r_c=r_c(x_{c,j})$ (middle part), and $\Delta s_j(x_{c,j})/
\Delta s_{j,{\rm ncr}}$ (lower part) for the proton with $\gamma=5/3$.
We set parameters as $\xi_{\rm cr}=0.5$ (purple line), $\xi_{\rm cr}=0.3$ (black line), and $\xi_{\rm cr}=0$ (green line)
with fixed values of $v_0=4000~{\rm km~s^{-1}}$, $T_0=3\times10^4$~K, and $1/\sqrt{\xi}_{\rm B}=v_0/(\delta B/\sqrt{4\pi\rho_0})=3$.
The entropy jump $\Delta s_{j,{\rm ncr}}$ for the case without the CRs (Model~0, and thus the case of the usual collisional shock)
is derived from
%
%%%%%%%%%%%%%%
\begin{eqnarray}
\left(\gamma-1\right)m_j\Delta s_{j,{\rm ncr}} = \ln x_{c,j,{\rm ncr}} - \gamma\ln r_{c,{\rm ncr}},
\label{eq:delta s ncr}
\end{eqnarray}
%%%%%%%%%%%%%%
%
where $x_{c,j,{\rm ncr}}$ and $r_{c,{\rm ncr}}$ are given by Model~0.
The right panel of figure~\ref{fig:fx} shows the sets of $\xi_{\rm cr}$, $x_{c,j}$, and $r_c$ satisfying $f=0$.
The function $f(x_{c,j})$ shows two solutions for a given
$\delta B$ depending on $P_{\rm cr}$. Although we don't have a precise explanation about these two solutions that may require a
full understanding of the ion heating by the kinetics theory, we may be able to interpret them from resultant downstream values.
Let us consider the case of $\xi_{\rm cr}=0$ in which $\Delta s_j/\Delta s_{j,{\rm ncr}}\approx1$ around each solution.
We will refer to the solution giving $x_{c,j}/{\cal M}_{{\rm s},j}{}^2\approx0.17$ and $r_c\approx1.27$ as `solution~A', while we will
refer to the other solution giving $x_{c,j}/{\cal M}_{{\rm s},j}{}^2\approx1.28$ and $r_c\approx8.31$ as `solution~B'.
The resultant temperature ($T_{j,2}/T_0=x_{c,j}/r_c\approx0.1 mv_0{}^2/\gamma kT_0$) is almost the same as each other.
This means that the speed of particles' random motion is almost the same as each solution. On the other hand, the difference in the
compression ratios indicates that the speed of particles' bulk motion is significantly different from each other. In a collisional shock
in the strong shock limit, the downstream temperature satisfies $(3/2)kT_2=m v'_{0}{}^2/2$, where $v'_0=v_0-v_2$ is the upstream
velocity measured in the downstream rest frame and we use $\gamma=5/3$. This might mean that since  our shock consumes its energy for the
generation of the nonthermal components, the random motion speed measured in the downstream rest frame  $\tilde{v}'_{\rm R}\equiv
\sqrt{3kT_{j,2}/m_j}$ should be equal or smaller than $v_0'=v_0-v_2$ for the solution representing the shock transition
(i.e. $\tilde{v}'_{\rm R}/v_0'\le1$). Solution~A gives the speed as $\tilde{v}'_{{\rm R}}/v_0'\simeq2.3$, while solution~B
gives $\tilde{v}'_{{\rm R}}/v_0'\simeq0.6$. Hence, solution~B may correspond to the shock transition.
Solution~A should be rejected because it does not satisfy the energy flux conservation law.
\par
When $\xi_{\rm cr}$ becomes large, the two solutions approach with each other, coinciding at $\xi_{\rm cr}\simeq0.3$ (multiple
roots), and finally, the solution vanishes. The multiple roots ($\xi_{\rm cr}=0.3$) give $\tilde{v}'_{\rm R}/v_0'\simeq0.7$ and
$\Delta s_j/\Delta s_{j,{\rm ncr}}\simeq0.93$.
Thus, the multiple roots may represent the shock transition giving the maximum $P_{\rm cr}$ feasible in our shock model.
In this article, we set the maximum $\xi_{\rm cr}$ to compare the no CR cases with the case of extremely efficient CR acceleration.
The maximum $\xi_{\rm cr}$ is derived from the multiple roots of $f=0$ with given $\xi_{\rm B}$.
\par
For the case of $v_0=4000~{\rm km~s^{-1}}$ and $T_0=3\times10^4$~K with given $1/\sqrt{\xi_{\rm B}}=3$, we obtain
the maximum acceptable CR production $\xi_{\rm cr}\simeq0.3$, $\Delta s_j/\Delta s_{j,{\rm ncr}}\simeq0.92\mathchar`-0.95$ depending on
$m_j$, $r_c\simeq3.29$, and $kT_{\rm p,2}\simeq14.4~{\rm keV}$.
Note that in the case of Model~0 (the usual collisional shock case), we obtain $r_c=4.00$ and
$kT_{\rm p,2}=31.3$~keV. The fraction of the CRs $\xi_{\rm cr}=0.3$ seems to be reasonable for the SNR shocks as sources of Galactic CRs.
From the subtraction of the energy fluxes of the thermal particles at the far upstream and downstream, we can regard that
roughly $50$~\% of the upstream energy flux is transferred to the nonthermal components. The fraction of magnetic
pressure $1/\sqrt{\xi_{\rm B}}=3$ corresponds to magnetic-field strength of $\delta B\simeq611~{\mu G}
 ( v_0 /4000~{\rm km~s^{-1}} ) ( n_{\rm p,0}/1~{\rm cm^{-3}})^{1/2} $ which is consistent with estimated strength from
X-ray observations of young SNRs~(e.g., \cite{vink03,bamba05,uchiyama07}). Thus, our parameter choice of $1/\sqrt{\xi_{\rm B}}=3$
can be reasonable to adopt our model to the young SNR shocks.
\par
Here we consider about the choice of the maximum $\xi_{\rm cr}$.
In the case of the collisional shock formed by the hard-sphere collisions, for example,
the collisions result in one of the most efficient randomizations of particles. Thus, the collisional shock can `easily' dissipate
its kinetic energy within the mean collision time. In the collisionless plasma, such efficient randomization process is absent.
The particles in the plasma tend to behave as `nonthermal' particles resulting in a generation of electromagnetic disturbances by themselves.
The collisionless shock is formed by the self-generated disturbances so that almost all particles become thermal particles.
Although the number of the nonthermal particles is very smaller than the number of the thermal particles, the efficient randomization
caused by the nonthermal particles is required to form the collisionless shock.
Our choice of the maximum $\xi_{\rm cr}$ corresponds that
the effect is minimized per one nonthermal particle.
\par
%
%%%%%%%%%%%%%%
\begin{figure}[htbp]
\begin{center}
\includegraphics[scale=0.65]{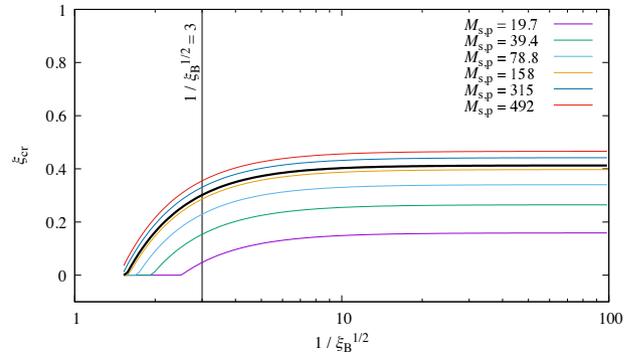}
\end{center}
\caption{The maximum $\xi_{\rm cr}$ derived from $f=0$
as a function of $1/\sqrt{ \xi_{\rm B} }=v_0/(\delta B/\sqrt{4\pi\rho_0})$ for given the shock velocity $v_0$
and $T_0=3\times10^4$~K. The heavy black solid line shows ${\cal M}_{\rm s,p}=197$ ($v_0=4000~{\rm km~s^{-1}}$).
The vertical thin line indicates $1/\sqrt{ \xi_{\rm B} }=3$.
For a comparison, we display
${\cal M}_{\rm s,p}=$
$19.7$ (purple, $v_0=  400~{\rm km~s^{-1}}$),
$39.4$ (green, $v_0=  800~{\rm km~s^{-1}}$),
$78.8$ (light blue, $v_0= 1600~{\rm km~s^{-1}}$),
$158 $ (orange, $v_0= 3200~{\rm km~s^{-1}}$),
$315 $ (blue, $v_0= 6400~{\rm km~s^{-1}}$), and
$492 $ (red, $v_0=10000~{\rm km~s^{-1}}$).}
\label{fig:Ma vs csi}
\end{figure}
%%%%%%%%%%%%%%
%
Figure~\ref{fig:Ma vs csi} shows the maximum $\xi_{\rm cr}$ derived from $f=0$ as a function of $1/\sqrt{\xi_{\rm B}}$ for
${\cal M}_{\rm s,p}=197$.
The fraction $\xi_{\rm cr}$ drops around $1/\sqrt{\xi_{\rm B}}\lesssim3$ but is flattened for $1/\sqrt{ \xi_{\rm B} }\gtrsim3$.
This depletion of the maximum $\xi_{\rm cr}$ is qualitatively obvious in terms of the energy budget of the shock; the upstream
kinetic  energy is divided into the thermal components, $P_{\rm cr}$ and $\delta B$.
The fraction of $\delta B$ is a given parameter. The fraction of the thermal components and the maximum fraction of $P_{\rm cr}$
are derived from the entropy production.
We will refer to this model with $1/\sqrt{\xi_{\rm B}}=3$ and the maximum $\xi_{\rm cr}$ as Model~3.
\par
%
%%%%%%%%%%%%%%
\begin{figure}[htbp]
\begin{center}
\includegraphics[scale=0.65]{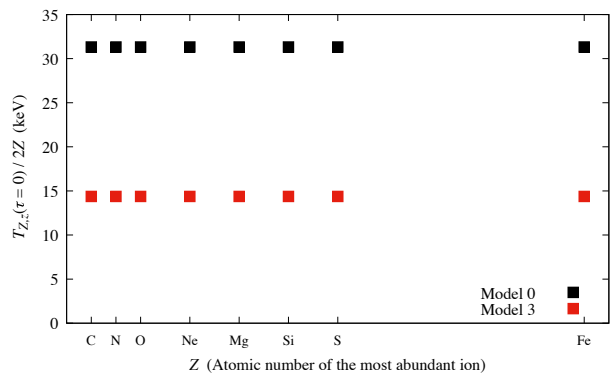}
\end{center}
\caption{The initial, downstream ion temperatures divided by $2Z$ (i.e., the particle mass in atomic mass unit)
for Model~0 and Model~3 with $v_0=4000~{\rm km~s^{-1}}$ and $T_0=3\times10^4$~K. The black square shows
the results of Model~0, and the red square shows the results of Model~3. The results
of Model~1 and Model~2 are the same as the results of Model~0, respectively. The horizontal axis shows the Atomic number $Z$.}
\label{fig:initial}
\end{figure}
%%%%%%%%%%%%%%
%
Figure~\ref{fig:initial} shows the results of downstream ion temperatures divided by $2Z$ (i.e., the particle mass
in atomic unit) for Model~0 and Model~3 with $v_0=4000~{\rm km~s^{-1}}$ and $T_0=3\times10^4$~K. The
reduced temperatures $kT_{Z,z}/2Z$ of Model~3 do not depend on the particle mass, indicating that the temperature
ratios between the ions are equal to their ion mass ratio.
Such mass-proportional ion temperatures are observed at SN~1987A~\citep{miceli19}.
The temperature jump $T_{j,2}/T_0$ is given by $x_{c,j}/r_c\sim{\cal M}_{{\rm s},j}$.
The relation of $x_{c,j}/r_c\sim{\cal M}_{{\rm s},j}$ is also implied by the condition of $f=0$.
Thus, Model~3 predicts that the ion temperature ratio is given by the mass ratio, similar to the case of Model~0.
Note that Model~1 and Model~2 give ion temperatures almost the same as Model~0. On the other hand, $kT_{Z,z}/2Z$ of Model~3
is smaller than the case of Model~0 by a factor of 2 due to the generations of the nonthermal components.
\par
The existence of more than two solutions is usually seen in the CR accelerating shock
model~(e.g., \cite{drury81,vink10,vink14}, and references therein). The unphysical solution like Solution~A of our
model, which does not satisfy the energy flux conservation law, also exists in previous studies.
The essential difference between our model and previous studies is the treatment of the randomization
in the shock transition.
In the previous studies, the randomization process, which determines the downstream thermal energy, may be implicitly chosen
to satisfy the flux conservation laws with assumed parameters ($P_{\rm cr}$, energy flux of escaping CRs, etc.).
\citet{vink14} also derived a critical sonic Mach number ${\cal M}_{\rm acc}=\sqrt{5}$ below which the particle
acceleration should not occur. In our model, a similar sonic Mach number may be derived from conditions of
$\Delta s_j\le\Delta s_{j,{\rm ncr}}$ and $T_{j,2}\le T_{j,2,{\rm ncr}}$, where
$T_{j,2,{\rm ncr}}$ is given by Model~0 (the equation \ref{eq:kT2 ncr}).
The former states that the generated entropy in the collisionless shock should be smaller than
in the case of collisional shocks. The latter states that the downstream temperature should be
smaller than the case of adiabatic, collisional shocks without the CRs.
Note that the entropy and temperature must be determined independently to derive the density or pressure
in thermodynamics. In other words,
the conditions $\Delta s_j\le\Delta s_{j,{\rm ncr}}$ and
$T_{j,2}\le T_{j,2,{\rm ncr}}$ are independent with each other.
From the equations (\ref{eq:delta s ncr}) and (\ref{eq:delta s}), and using the relation of
$m_jv_0{}^2/kT_0=(\rho_{j,0}/\rho_0)\gamma{\cal M}_{\rm s}{}^2$, we can derive
%
%%%%%%%%%%%%%%%%%%%%%%
\begin{eqnarray}
\frac{ T_{j,2,{\rm ncr}} }{ T_0 }
\ge
\frac{ T_{j,2}           }{ T_0 }
\ge
\frac{\rho_{j,0}               }{\rho_0}
\frac{ \gamma(\gamma-1) {\cal M}_{\rm s}{}^2 \left( \langle\tilde{v}_j\rangle/v_0 \right)^2 }
     { \ln{\left(T_{j,2,{\rm ncr}}/T_0\right)} - \left(\gamma-1\right)\ln r_{ c,{\rm ncr}} },
\label{eq:ineq}
\end{eqnarray}
%%%%%%%%%%%%%%%%%%%%%%
%
where
%
%%%%%%%%%%%%%%%%%%%%%%
\begin{eqnarray}
\frac{ T_{j,2,{\rm ncr}} }{ T_0 }
=\frac{ \rho_{j,0} }{ \rho_0 }
 \frac{ \gamma{\cal M}_{\rm s}{}^2 }{ r_{c,{\rm ncr}} }
 \left( 1 - \frac{ 1 }{ r_{c,{\rm ncr}} } + \frac{1}{\gamma{\cal M}_{\rm s}{}^2 } \right),
\end{eqnarray}
%%%%%%%%%%%%%%%%%%%%%%
%
and $r_{c,{\rm ncr}}$ is given by the equation~(\ref{eq:rc ncr}).
The critical Mach number ${\cal M}_{\rm s,acc}$ is given when the equal sign of the inequality holds.
Regarding $\langle\tilde{v}_j\rangle\simeq v_0$ and $\rho_j/\rho_0\simeq1$ for simplicity,
we obtain the numerical value of ${\cal M}_{\rm s,acc}\simeq16.34$
above which we can find sets of $\xi_{\rm cr}$ and $\xi_{\rm B}$ satisfying the inequality~(\ref{eq:ineq}).
The larger critical Mach number than that derived by \citet{vink14} may be due to the difference in the assumed
randomization process. However, the value of ${\cal M}_{\rm s,acc}$
may also depend on the Alfv{\'e}n Mach number, whose effects are not studied in this paper.
When the sonic Mach number decreases due to a shock deceleration, the effects of the mean magnetic
field at the far upstream region can be important.
The shocks with a lower Mach number are seen in the solar wind at coronal mass ejection events, clusters of galaxy, and so on.
In predictions of the accelerated paritcles amount in such cases, we shoud include the pre-existing ordered magnetic field to
the flux conservation laws and evaluation of $\bm{J}\cdot\bm{E}$ term, differing from the current approach.
We will study a general critical Mach number with more elaborate models in future work.
\par
Finally, we parameterize the electron heating for the case of the extremely efficient CR acceleration as
%
%%%%%%%%%%%%%%%
\begin{eqnarray}
&& \varepsilon_{\rm p,2} = \varepsilon_{{\rm p,2,Model3}}(1-f_{\rm eq}), \nonumber \\
&& \varepsilon_{\rm e,2} = \varepsilon_{{\rm e,2,Model3}}+\varepsilon_{\rm p,2,Model3}f_{\rm eq},
\end{eqnarray}
%%%%%%%%%%%%%%%
%
where $\varepsilon_{\rm p,2,Model3}$ and $\varepsilon_{\rm e,2,Model3}$ are the internal energy
calculated by Model~3. Here we have supposed an additional energy transfer: the internal energy of the thermal protons
is transferred to the thermal electrons. The fraction of the transferred internal energy is written by the temperature
ratio of $\beta=T_{\rm e,2}/T_{\rm p,2}$ as
%
%%%%%%%%%%%%%%%
\begin{eqnarray}
f_{\rm eq} = \frac{ \beta(n_{\rm e,0}/n_{\rm p,0})-(\varepsilon_{\rm e,2,Model3}/\varepsilon_{\rm p,2,Model3}) }
                  { \beta(n_{\rm e,0}/n_{\rm p,0}) + 1 }.
\end{eqnarray}
%%%%%%%%%%%%%%%
%
\citet{laming14} pointed that the electron temperature can be significantly large ($\beta\sim0.1$)
when the shock accelerates the CRs efficiently.  We calculate the cases of $\beta=0.01$ (Model~4)
and $\beta=0.1$ (Model~5) in this paper. Table~\ref{tab:model summary} shows a summary of our shock models.
%
%%%%%%%%%%%%%%
\begin{table}[htbp]
\tbl{Summary of the shock jump models. We set $v_0=4000~{\rm km~s^{-1}}$
and $T_0=3\times10^4$~K.  From the left-hand side to the right-hand side, the columns indicate
the model name, the compression ratio $r_c$, the downstream proton temperature $kT_{\rm p,2}$, the downstream electron temperature
$kT_{\rm e,2}$, the fraction of the amplified magnetic field $\xi_B=\delta B^2/(4\pi\rho_0 v_0{}^2)$, and the fraction of the CR
pressure $\xi_{\rm cr}=P_{\rm cr}/(\rho_0 v_0{}^2)$.}{%
\begin{tabular}{cc cc cc}
\hline
Model & $r_c$ & $kT_{\rm p,2}$ & $kT_{\rm e,2}$ ($\beta=T_{\rm e,2}/T_{\rm p,2}$) & $\xi_{\rm B}$ & $\xi_{\rm cr}$  \\
\hline
0     &   4   & 31.32~keV      & 17.1~eV  ($m_{\rm e}/m_{\rm p}$)        & 0             & 0              \\
1     &   4   & 31.01~keV      & 31.0~eV  (0.01)                         & 0             & 0              \\
2     &   4   & 28.34~keV      & 2.83~keV (0.1)                          & 0             & 0              \\
3     & 3.29  & 14.38~keV      & 8.62~eV  (1.1$m_{\rm e}/m_{\rm p}$)     & 1/9           & 0.30           \\       
4     & 3.29  & 14.23~keV      & 14.2~eV  (0.01)                         & 1/9           & 0.30           \\
5     & 3.29  & 13.01~keV      & 1.30~keV (0.1)                          & 1/9           & 0.30           \\
\hline                                
\end{tabular}}
\label{tab:model summary}
\end{table}
%%%%%%%%%%%%%%
%

\section{evolution track of the downstream ionization balance and temperatures}
\label{sec:results}
%
%%%%%%%%%%%%%%
\begin{figure}[htbp]
\begin{center}
\includegraphics[scale=0.65]{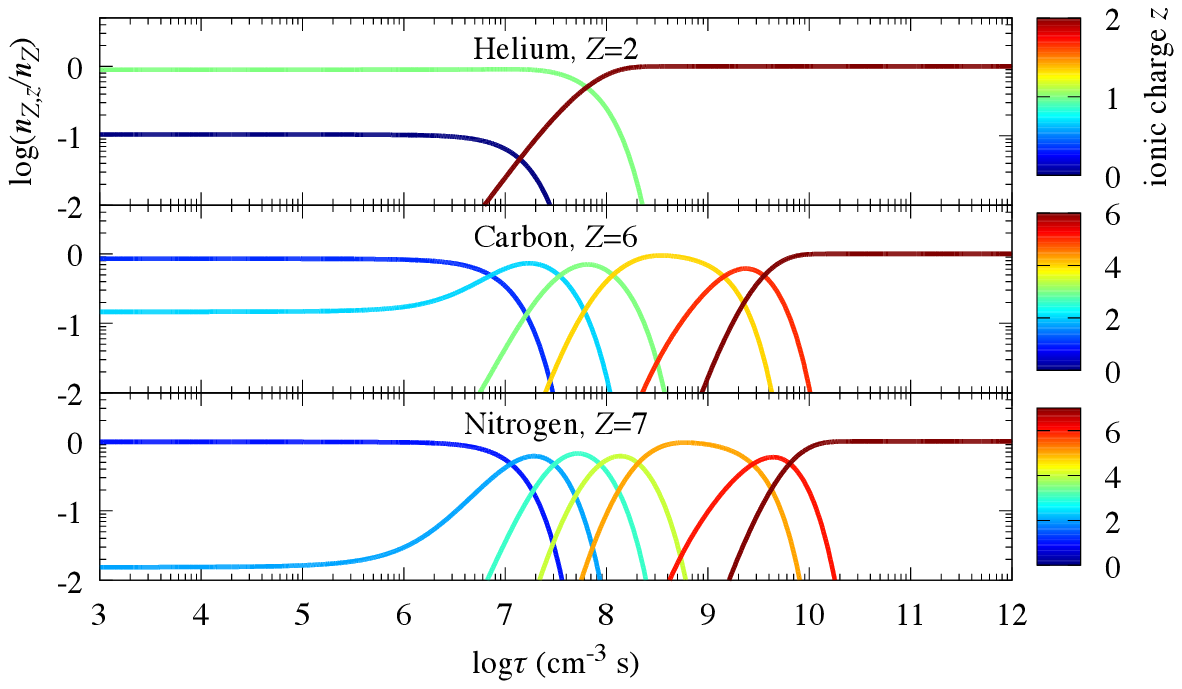}
\includegraphics[scale=0.65]{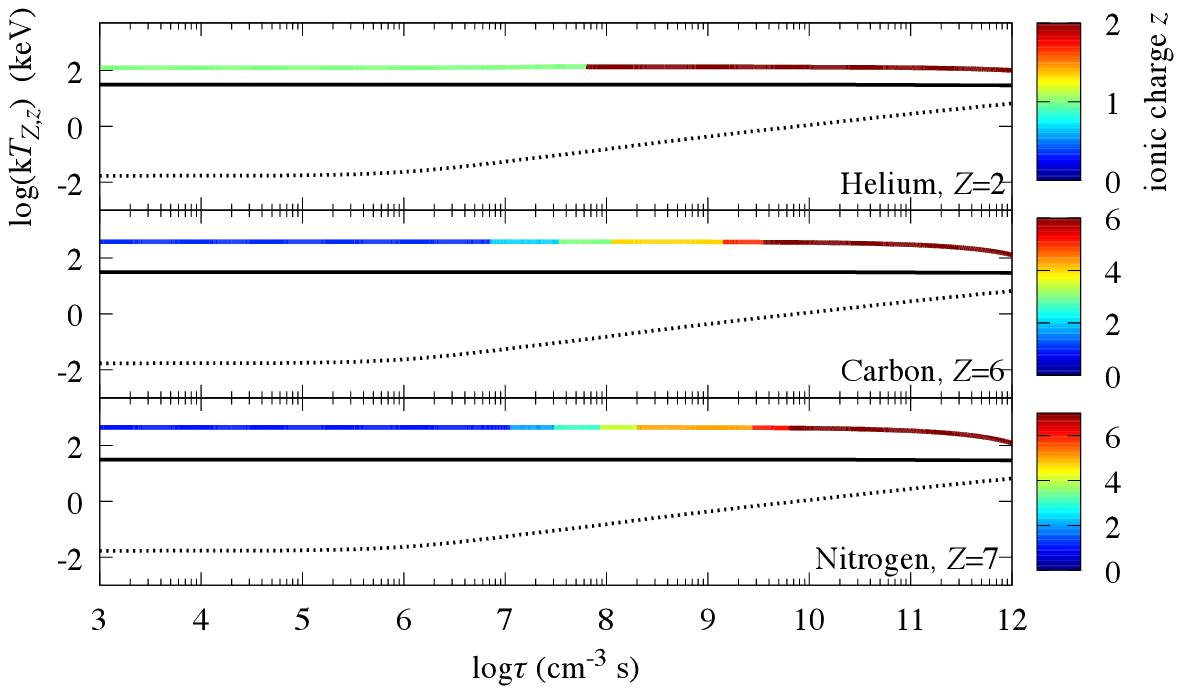}
\end{center}
\caption{The ionization balance $n_{Z,z}/n_Z$ (left panel) and temperature $kT_{Z,z}$ of
the most abundant species among its ionic charge (right panel)
for Model~0 with the shock velocity of $v_0=4000~{\rm km~s^{-1}}$.
We display the species He, C, and N. The color represents the ionic charge $z$ of ion. The solid black line
and black dots are the temperatures of proton and electron, respectively.}
\label{fig:light}
\end{figure}
%%%%%%%%%%%%%%
%
%%%%%%%%%%%%%%
\begin{figure}[htbp]
\begin{center}
\includegraphics[scale=0.65]{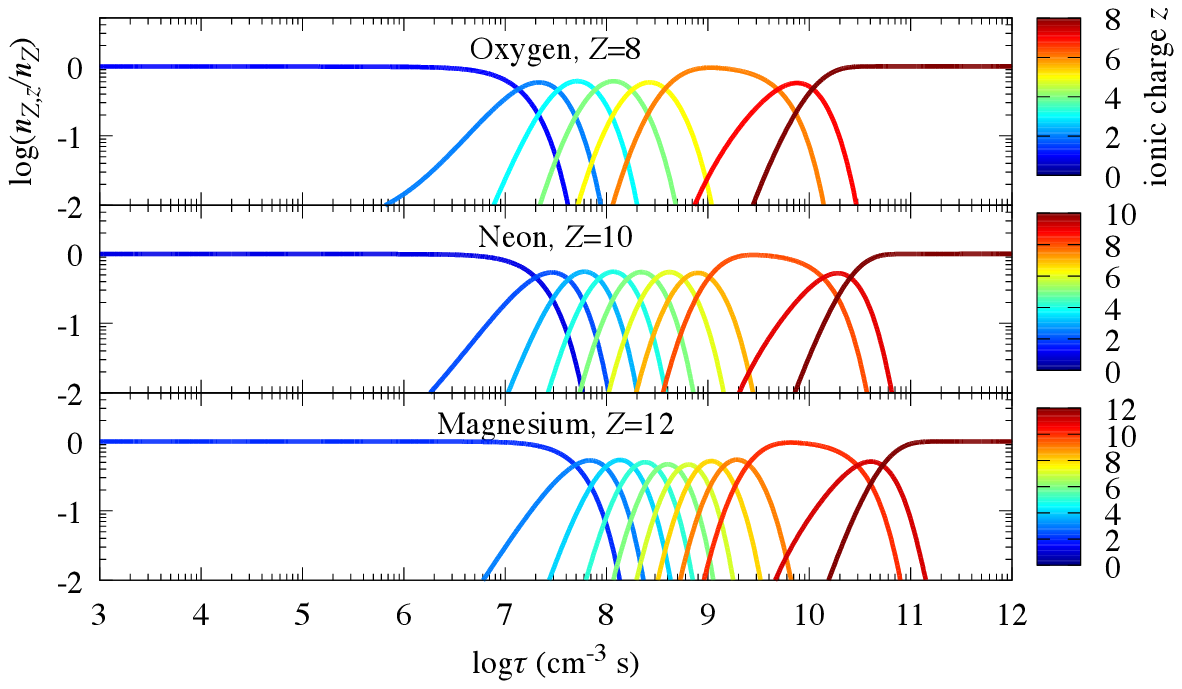}
\includegraphics[scale=0.65]{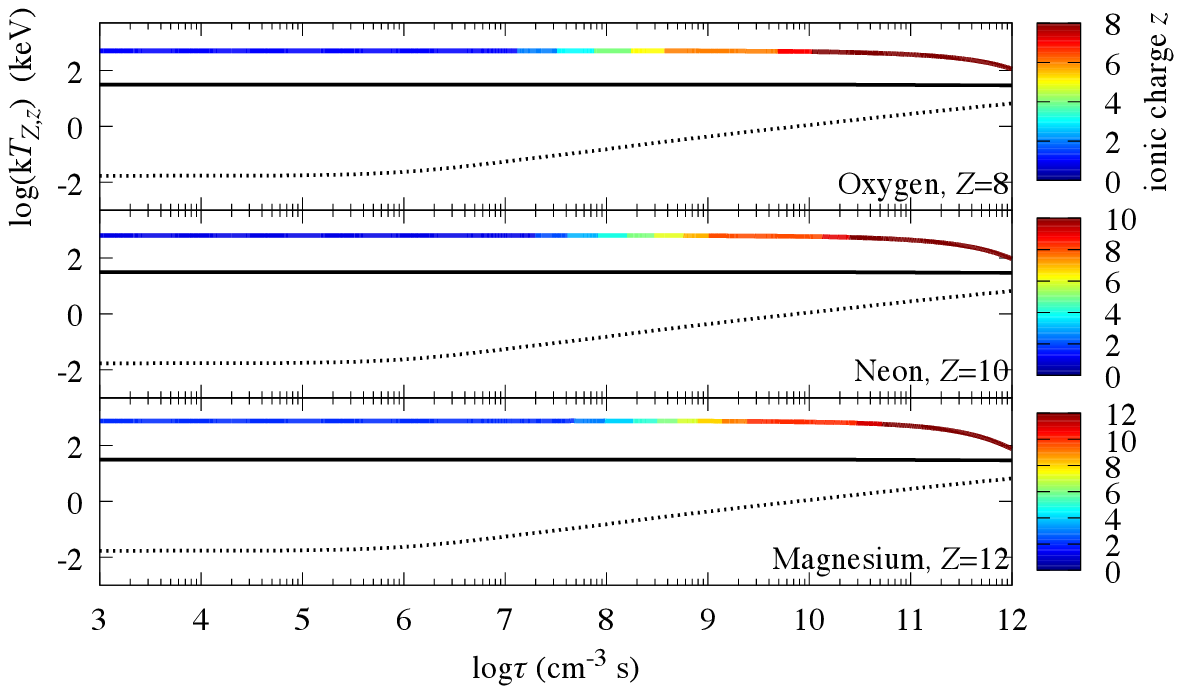}
\end{center}
\caption{The ionization balance $n_{Z,z}/n_Z$ (left panel) and temperature $kT_{Z,z}$ of
the most abundant species among its ionic charge (right panel)
for Model~0 with the shock velocity of $v_0=4000~{\rm km~s^{-1}}$.
We display the species O, Ne, and Mg. The color represents the ionic charge $z$ of ion. The solid black line
and black dots are the temperatures of proton and electron, respectively.}
\label{fig:medium}
\end{figure}
%%%%%%%%%%%%%%
%
%
%%%%%%%%%%%%%%
\begin{figure}[htbp]
\begin{center}
\includegraphics[scale=0.65]{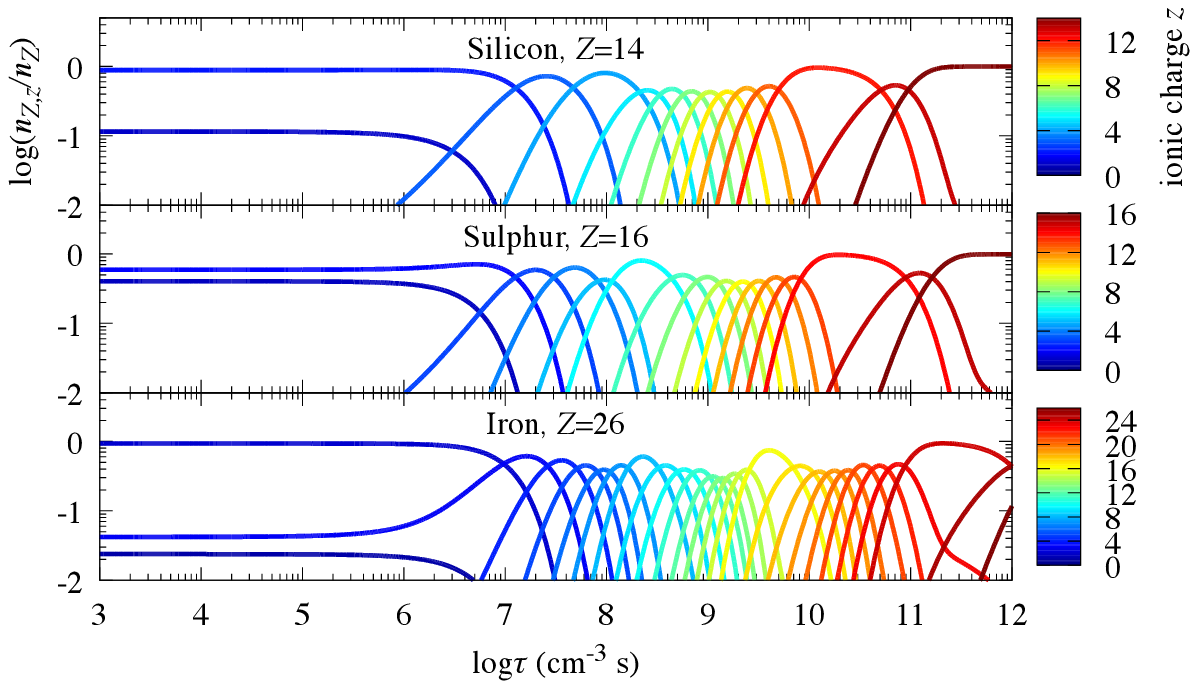}
\includegraphics[scale=0.65]{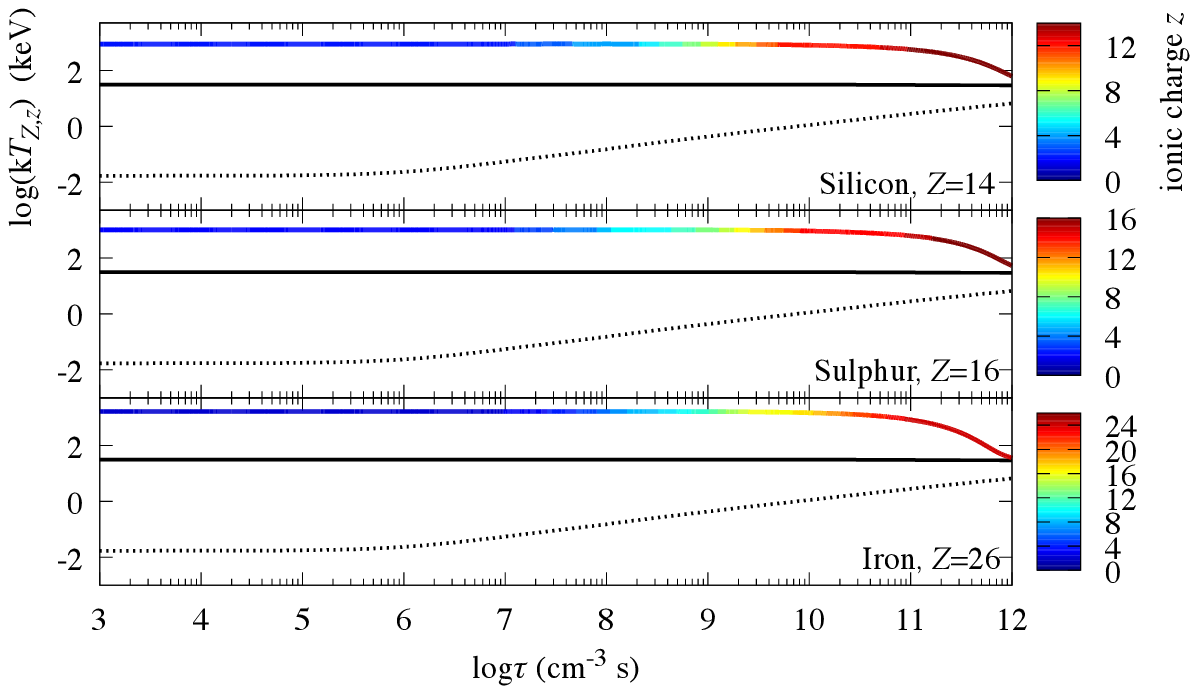}
\end{center}
\caption{The ionization balance $n_{Z,z}/n_Z$ (left panel) and temperature $kT_{Z,z}$ of
the most abundant species among its ionic charge (right panel)
for Model~0 with the shock velocity of $v_0=4000~{\rm km~s^{-1}}$.
We display the species Si, S, and Fe. The color represents the ionic charge $z$ of ion. The solid black line
and black dots are the temperatures of proton and electron, respectively.}
\label{fig:heavy}
\end{figure}
%%%%%%%%%%%%%%
%
Here we show the results of the ionization balance and temperature relaxation in the downstream region,
omitting the effects of the expansion ($dV/dt=0$) as a reference.
For convenience, we introduce $d\tau=ndt$, where $n$ is the total number density, so that
%
%%%%%%%%%%%%%%
\begin{eqnarray}
\frac{d\varepsilon_j}{d\tau}
&\approx& \frac{\dot{q}_j}{n}, \\
\frac{dn_{Z,z}}{d\tau}
&=& \frac{ n_{\rm e} }{ n }
\left[   R_{Z,z-1} n_{Z,z-1}
-  \left( R_{Z,z} + K_{Z,z} \right) n_{Z,z}
\right. \nonumber \\
&+&\left.
+K_{Z,z+1} n_{Z,z+1}
\right],
\end{eqnarray}
%%%%%%%%%%%%%%
%
where we use $\rho=$const.
Figures~\ref{fig:light} (for He, C, and N), \ref{fig:medium} (for O, Ne, and Mg), and \ref{fig:heavy} (for Si, S, and Fe)
show $n_{Z,z}/n_Z$ and $kT_{Z,z}$ for Model~0 with the shock velocity of $v_0=4000~{\rm km~s^{-1}}$, where
$n_Z=\sum_{z}^{}n_{Z,z}$ is the total number density of the atoms with the atomic number $Z$. Note that $\tau=\int ndt
\simeq nt$ because of the small neutral fraction. Here we display the ion temperatures for
the most abundant species among its ionic charge.
\par
The evolution tracks of $n_{Z,z}/n_Z$ and $T_{Z,z}$ for other models are not so different from the case of Model~0.
In the case of a higher electron temperature (Model~1 and Model~2), the ions are quickly ionized.
%
%%%%%%%%%%%%%
\begin{figure}
\begin{center}
\includegraphics[scale=0.65]{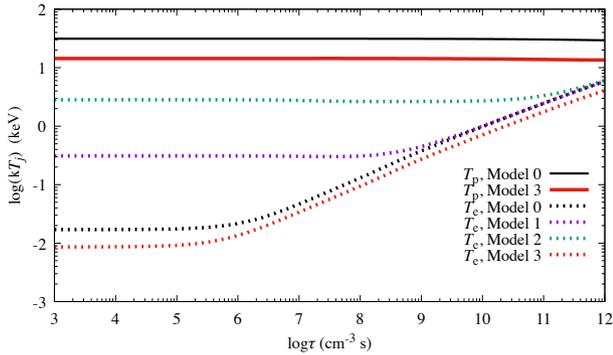}
\end{center}
\caption{The electron temperatures for Model~0 (black dots), 1 (purple dots), and 2 (green dots)
with $v_0=4000~{\rm km~s^{-1}}$. The solid black line shows the proton temperature for Model~0.
The red solid line and red dots are the proton and electron temperatures of Model~3, respectively.}
\label{fig:Te}
\end{figure}
%%%%%%%%%%%%%
%
Figure~\ref{fig:Te} shows the electron temperatures for Model~0, 1, 2, and 3 with $v_0=4000~{\rm km~s^{-1}}$.
The relation between Model~4 and Mode~1 (Model~5 and Model~2) is similar to that of Model~3 and Model~0.
The ionization balance $n_{Z,z}/n_Z$ becomes the same in each model after the electron temperature coincides.
Note that the electron temperature increases within a column density scale of $ntV_{\rm sh}\sim10^{14}~{\rm cm^{-2}}
(nt/10^6~{\rm cm^{-3}~s})(V_{\rm sh}/4000~{\rm km~s^{-1}})$. This column density scale is comparable to the size
of the H$\alpha$ emission region~(e.g., \cite{shimoda19a}). Therefore, to study the electron heating at the shock, the H$\alpha$
observation may be better than the X-ray line observations.
In the case of a lower ion temperature due to the production of $P_{\rm cr}$ and $\delta B$ (Model~3), the temperature equilibrium
is achieved at a smaller $nt$ (e.g., the temperature of Fe is equal to the proton's at $nt\simeq2\times10^{11} ~{\rm cm^{-3}~s}$)
because the relaxation time of the Coulomb collision depends on $T^{3/2}$~\citep{spitzer62}. Note that the lower electron temperature
results in a lower ionization state at given $\tau\simeq nt$.
\par
%
%%%%%%%%%%%%%%
\begin{figure}[htbp]
\begin{center}
\includegraphics[scale=0.65]{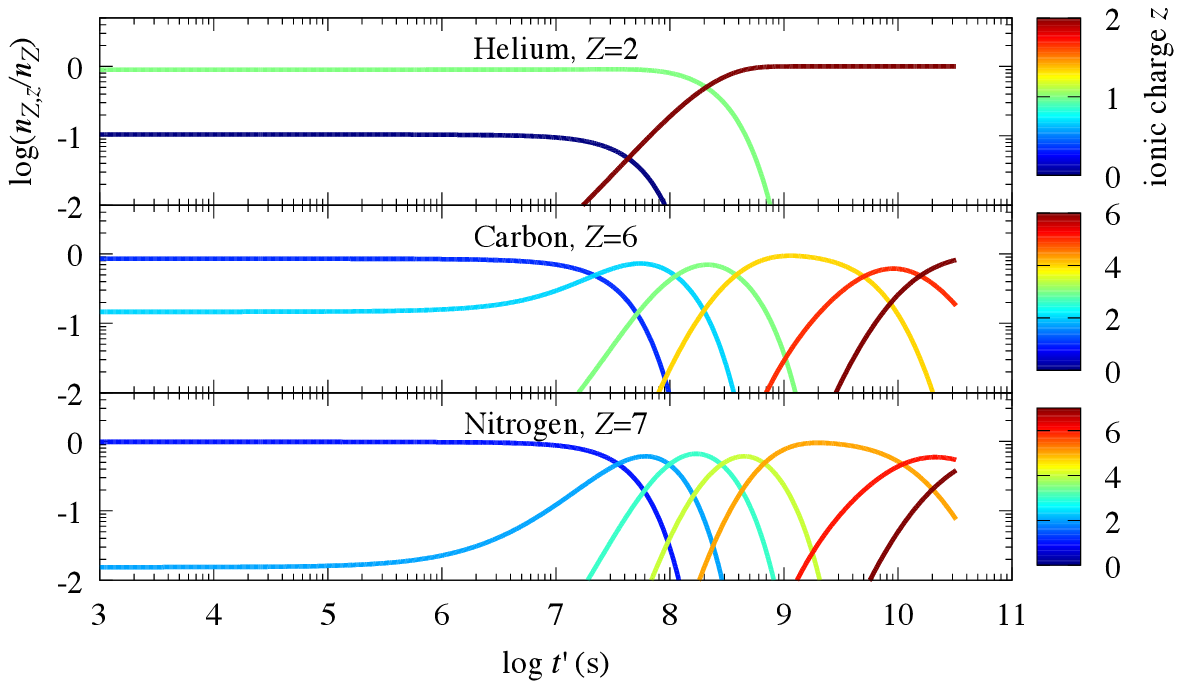}
\includegraphics[scale=0.65]{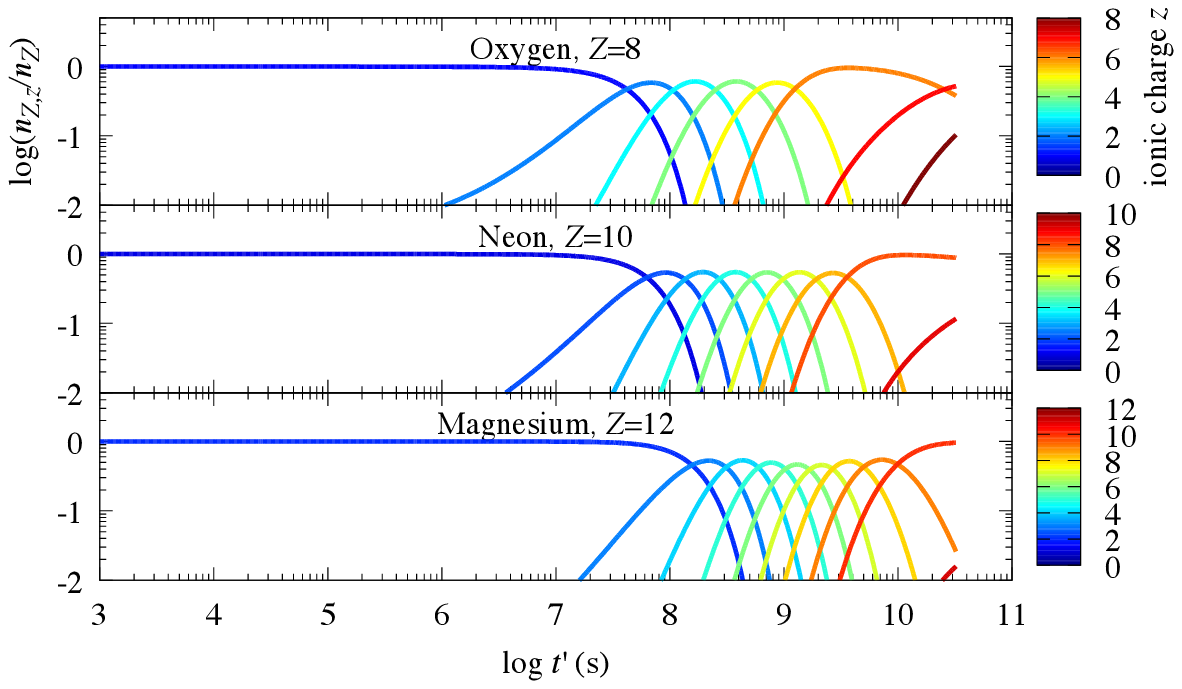}
\includegraphics[scale=0.65]{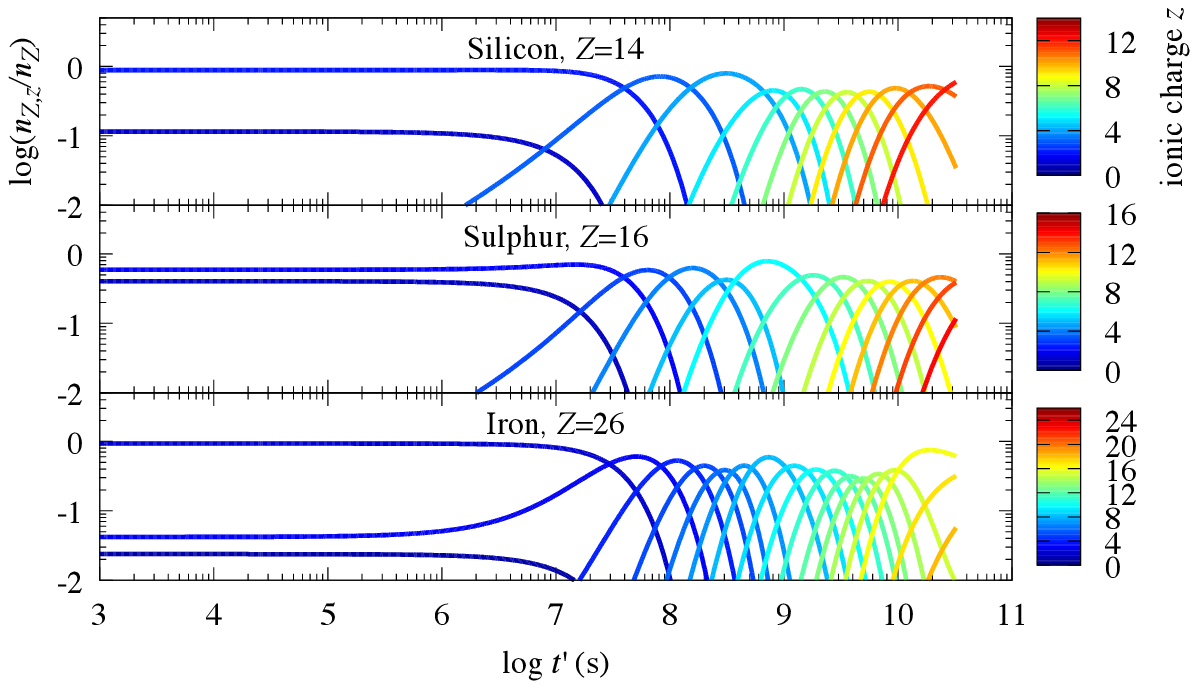}
\end{center}
\caption{The ionization balance $n_{Z,z}/n_Z$ for Model~0 with
$\rho_0=4.08\times10^{-2}m_{\rm p}$, $t_{\rm age}=1836~{\rm yr}$, $V_{\rm sh}(t_{\rm age})=3000~{\rm km~s^{-1}}$,
and $r/R_{\rm sh}(t_{\rm age})=0.8$.
We display the species He, C, N (top left panels), O, Ne, Mg (top right panels), Si, S, and Fe (bottom panels).
The color represents the ionic charge $z$ of the ion. The horizontal axis shows the time $t'=t-t_*$, where
$t_*$ is the shock transition time of the fluid parcel currently at $r/R_{\rm sh}(t_{\rm age})=0.8$.}
\label{fig:abundance expansion}
\end{figure}
%%%%%%%%%%%%%%
%
%%%%%%%%%%%%%
\begin{figure}
\begin{center}
\includegraphics[scale=0.65]{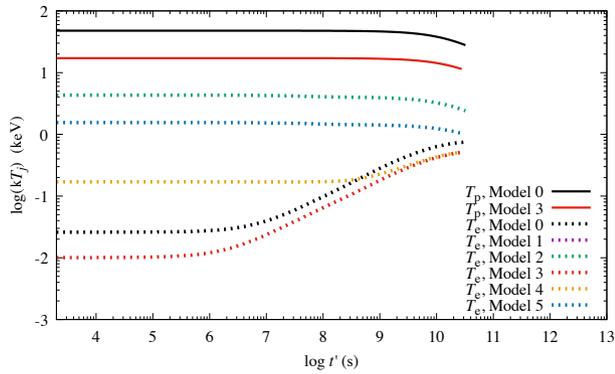}
\end{center}
\caption{The electron temperatures for Model~0 (black dots), 1 (purple dots), 2 (green dots), 3 (red dots),
4 (orange dots), and 5 (blue dots) with
$\rho_0=4.08\times10^{-2}m_{\rm p}$, $t_{\rm age}=1836~{\rm yr}$, $V_{\rm sh}(t_{\rm age})=3000~{\rm km~s^{-1}}$,
and $r/R_{\rm sh}(t_{\rm age})=0.8$. The solid black (red) line shows the proton temperature for Model~0
(Model~3). The proton temperatures for Model~1
and Model~2 (Model~4 and Model~5) are the almost same as Model~0 (Model~3).
}
\label{fig:Te expansion}
\end{figure}
%%%%%%%%%%%%%
%
When the effects of the expansion become important, we cannot characterize
the evolution only by $\tau\simeq nt$ and we should introduce parameters to describe the expansion of SNRs
and the observed position $r/R_{\rm sh}(t_{\rm age})$. Here we set $\rho_0=4.08\times10^{-2}m_{\rm p}$, $t_{\rm age}=1836~{\rm yr}$, and
$V_{\rm sh}(t_{\rm age})=3000~{\rm km~s^{-1}}$ for example. This parameter set will be used in comparisons of our Model to
the SNR RCW~86 (discuss later in Sect.~\ref{sec:synthetic}). Figure~\ref{fig:abundance expansion} shows the downstream
ionization structure of He, C, N (top panels), O, Ne, Mg (middle panels), Si, S, and Fe (bottom panels). The fluid parcel currently at
$r/R_{\rm sh}(t_{\rm age})=0.8$ crossed the shock at the time of $t_*$ when the shock velocity was
$V_{\rm sh}(t_*)=5094~{\rm km~s^{-1}}$ for Model~0, 1, and 2 ($4565~{\rm km~s^{-1}}$ for Model~3, 4, and 5).
Since the compression ratio depends on whether the CRs exist, the shock transition time $t_*$, shock velocity
$V_{\rm sh}(t_*)$, and $T_{\rm e}(t_*)$ are different from each model for the fluid parcel currently at $r/R_{\rm sh}(t_{\rm age})$.
The evolution of the ionization balance is similar to the case of the plane-parallel shock
until $t'\sim10^9\mathchar`-10^{10}~{\rm s}\sim t_{\rm age}$.
The cooling due to the expansion becomes important at $t'\sim t_{\rm age}$.
The ion temperatures decrease before the ions are well ionized due to
the expansion (decreasing of the density, ion temperature, and electron temperature). Figure~\ref{fig:Te expansion} shows the electron
temperatures for Model~0 (black dots), 1 (purple dots), 2 (green dots), 3 (red dots), 4 (orange dots), and 5 (blue dots).

\section{Synthetic Observations}
\label{sec:synthetic}
In this section, we perform synthetic observations of the shocked plasma considering the effects of turbulence
for the case of the SNR RCW~86. Since we do not calculate the overall spectrum of the emitted photons which needs
enormous calculations about emission lines, we mainly estimate the line shape.
\par
The SNR RCW~86 is one of the best targets for the study of the CR injection via the ion temperatures 
because the shells of the SNR show different thermal/nonthermal features from position to position
\citep{bamba00,borkowski01,tsubone17}.
The RCW~86 is considered as a historical SNR of SN~185~\citep{vink06}. Thus, we set $t_{\rm age}=1836~{\rm yr}$.
Along the northeastern shell of the RCW~86, the dominant X-ray radiation changes from thermal to
synchrotron emission~\citep{vink06}.
The thermal emission-dominated region is referred to `E-bright' region, and the synchrotron one is referred
to `NE' region.
The ionization age at NE is estimated as $\tau=(2.25\pm0.15)\times10^9~{\rm cm^{-3}~s}$ though this estimate
potentially contains errors due to the lack of the thermal continuum emissions~\citep{vink06}.
The E-bright region is fitted by two plasma components: (i) $\tau=(6.7\pm0.6)\times10^{9}~{\rm cm^{-3}~s}$, and
(ii) $\tau=(17\pm0.5)\times10^9~{\rm cm^{-3}~s}$. Both E-bright and NE show clear
O\emissiontype{VII}~He$\alpha$ and Ne\emissiontype{IX}~He$\alpha$ line emissions.
From the width of the synchrotron-emitting region (NE), the magnetic-field strength
is estimated as $\approx24\pm5~{\rm \mu G}$~\citep{vink06}.
\par
\citet{yamaguchi16} measured proper motions around these regions (not exactly the
same regions) as $v_0=720\pm360~{\rm km~s^{-1}}$ (E-bright), $v_0=1780\pm240~{\rm km~s^{-1}}$ (upper part of NE referred to `NE$_b$'), and
$v_0=3000\pm340~{\rm km~s^{-1}}$ (lower part of NE referred to `NE$_f$'). In the case of Model~3, the fractions of CR pressure $\xi_{\rm cr}$ become
$\xi_{\rm cr,720}\simeq0.14$ for $v_0=720~{\rm km~s^{-1}}$,
$\xi_{\rm cr,1780}\simeq0.24$ for $v_0=1780~{\rm km~s^{-1}}$, and
$\xi_{\rm cr,3000}\simeq0.28$ for $v_0=3000~{\rm km~s^{-1}}$, respectively.
If we simply suppose $\rho_0=(\tau/t_{\rm age})m_{\rm p}$ and adopt $1/\sqrt{\xi_{\rm B}}=3$, the CR pressure
and $\delta B$ of each region becomes
$P_{\rm cr, 720}\sim0.2~{\rm keV~cm^{-3}}$ and $\delta B_{\rm  720}\sim51~{\rm \mu G}$,
$P_{\rm cr,1780}\sim0.3~{\rm keV~cm^{-3}}$ and $\delta B_{\rm 1780}\sim55~{\rm \mu G}$, and
$P_{\rm cr,3000}\sim1.1~{\rm keV~cm^{-3}}$ and $\delta B_{\rm 3000}\sim93~{\rm \mu G}$,
where we adopt $\tau=12.0\times10^{9}~{\rm cm^{-3}~s}$ for the E-bright region as an average of the two components and
$\tau=2.25\times10^{ 9}~{\rm cm^{-3}~s}$ for the NE region, respectively.
If we adopt $v_0=360~{\rm km~s^{-1}}$ for the E-bright region, we obtain $\xi_{\rm cr,360}\sim2.9\times10^{-2}$.
The thermal-dominated E-bright region results from the higher density than the density at the NE region.
The magnetic-field strength $\delta B$ is the almost same as one another. 
Note that \citet{vink06} estimated the electron density at the E-bright region as $\sim0.6\mathchar`-1.5~{\rm cm^{-3}}$
from the emission measure with assuming the volume of the emission region. Our model predicts the downstream density
as $r_c\rho_0/m_{\rm p}\approx0.55~{\rm cm^{-3}}$ for the E-bright region with $v_0=720~{\rm km~s^{-1}}$
that is consistent with the previous estimate.
For the NE region, the number density is not well constrained because of the lack of the thermal continuum component.
Thus, our choice of model parameters can be
consistent with the observations of the RCW~86. In the following, we apply our model to the NE region
setting the parameters as
$t_{\rm age}=1836~{\rm yr}$,
$V_{\rm sh}(t_{\rm age})=3000~{\rm km~s^{-1}}$, and
$\rho_0/m_{\rm p}=\tau/t_{\rm age}=4.08\times10^{-2}~{\rm cm^{-3}}$, where $\tau=2.25\times10^9~{\rm cm^{-3}~s}$ is used.
We suppose that the downstream region from $r=R_{\rm obs}=0.8R_{\rm sh}(t_{\rm age})$ to $r=R_{\rm sh}(t_{\rm age})$ is observed.
Then, our model supposes that the expansion follows the Sedov-Taylor model during a time of $\Delta t\ge t_{\rm age}-t_*(R_{\rm obs})
= \left\{ 1 - (R_{\rm obs}/R_{\rm sh})^{r_c} \right\} t_{\rm age}\approx 0.6 t_{\rm age}$, where $r_c=4$ and the equation~(\ref{eq:t_*}) are used.
If the RCW~86 expands with a velocity of $\sim10^9~{\rm cm~s^{-1}}$ on average before entering the Sedov-Taylor stage, we
effectively assume the radius at the transition time of $t_0\approx0.6t_{\rm age}$ as $R_0\sim10^9~{\rm cm~s^{-1}}\times0.6t_{\rm age}\sim11$~pc.
Then, the radius at the current time is $R_{\rm sh}(t_{\rm age})\sim R_0(1/0.6)^{2/5}\sim13.5$~pc
which can be consistent with the actual radius of $\sim15$~pc (the distance is assumed as $2.5$~kpc).
\par
%
%%%%%%%%%%%%%
\begin{figure}
\begin{center}
\includegraphics[scale=0.65]{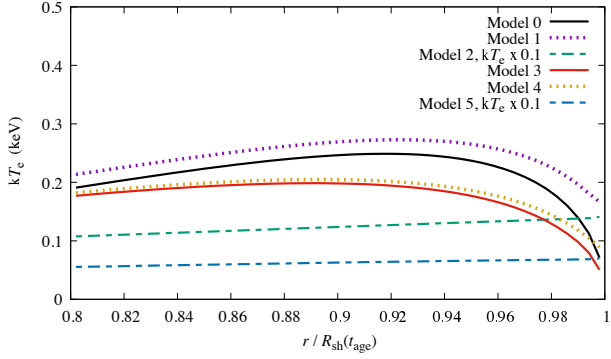}
\end{center}
\caption{
The radial profile of the electron temperature at $t_{\rm age}=1826~{\rm yr}$
for the NE region of RCW~86 with $V_{\rm sh}(t_{\rm age})=3000~{\rm km~s^{-1}}$ and
$\rho_0/m_{\rm p}=1.29\times10^{-2}$.
We display the results of Model~0 (black solid line), Model~1 (purple dots),
Model~2 (green broken line), Model~3 (red solid line), Model~4 (orange dots),
and Model~5 (blue broken line).}
\label{fig:Te rcw86}
\end{figure}
%%%%%%%%%%%%%
%
%
%%%%%%%%%%%%%
\begin{figure}
\begin{center}
\includegraphics[scale=0.65]{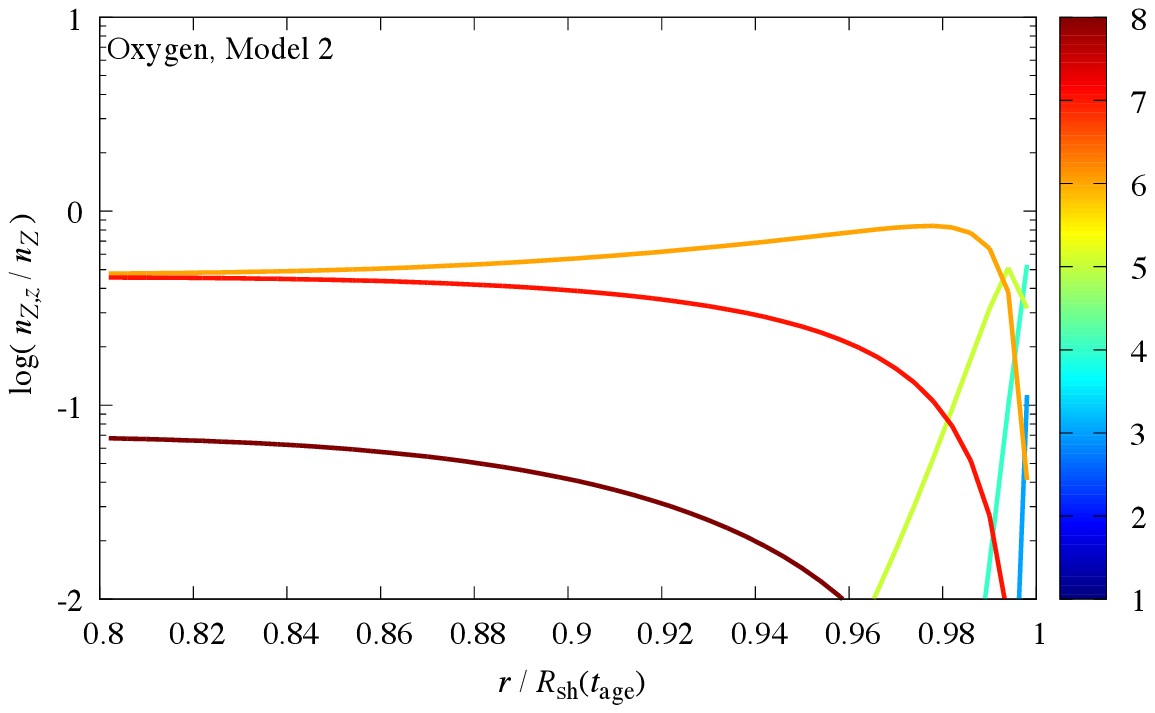}
\includegraphics[scale=0.65]{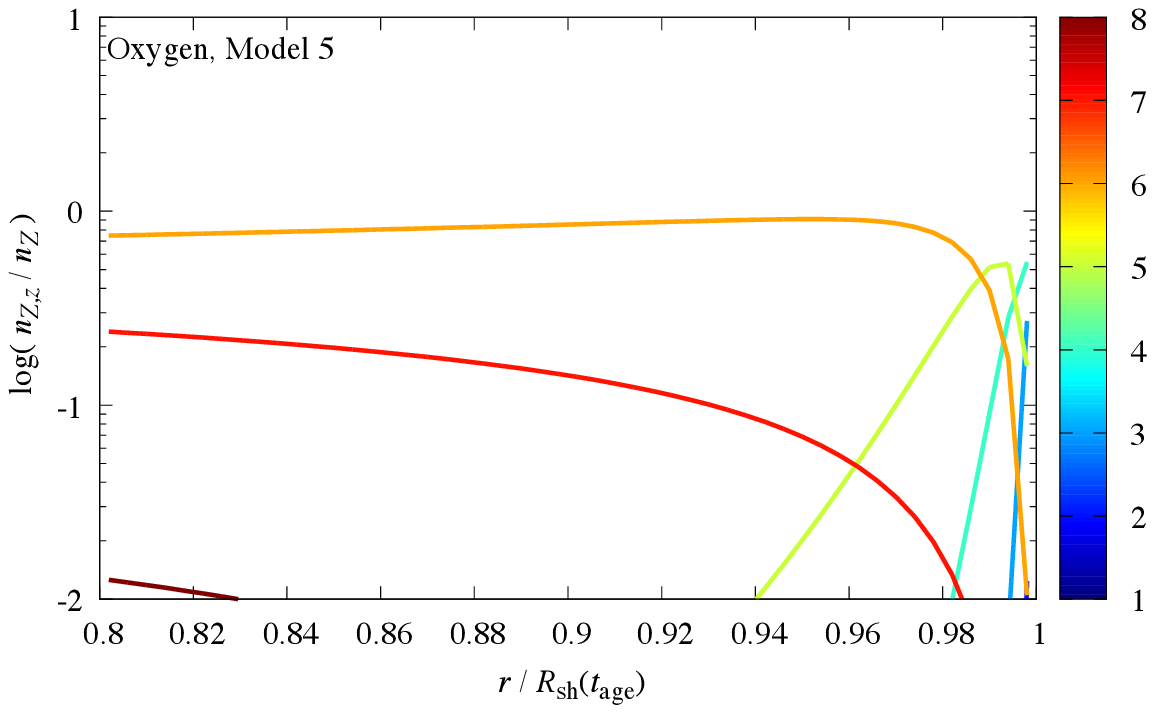}
\end{center}
\caption{
The radial profile of the oxygen abundance $n_{Z,z}/n_Z$ at $t_{\rm age}=1826~{\rm yr}$
for the NE region of RCW~86 with $V_{\rm sh}(t_{\rm age})=3000~{\rm km~s^{-1}}$ and
$\rho_0/m_{\rm p}=1.29\times10^{-2}$. The left panel shows the results of Model~2 and
the right panel shows Model~5. The color indicates the ionic charge $z$ of the ion.}
\label{fig:oxygen abundance rcw86}
\end{figure}
%%%%%%%%%%%%%
%
Figure~\ref{fig:Te rcw86} shows the radial profile of the electron temperature at $t=t_{\rm age}$
for Model~0 (black solid line), Model~1 (purple dots),
Model~2 (green broken line), Model~3 (red solid line), Model~4 (orange dots),
and Model~5 (blue broken line). To reproduce the bright O\emissiontype{VII}~He$\alpha$,
a relatively high electron temperature is preferred in terms of the excitation~($\sim1$~keV, see also \cite{vink06}),
though it is degenerating by the number density uncertainty.
Note that the excitation rate is $C_{l,u}\propto\exp(-E_{ul}/kT_{\rm e})/\sqrt{T_{\rm e}}$ and $E_{u,l}\simeq0.574$~keV
for O\emissiontype{VII}~He$\alpha$.
Thus, we mainly consider Model~2 ($\beta=T_{\rm e,2}/T_{\rm p,2}=0.1$ without the CRs) and Model~5 ($\beta=0.1$ with the CRs).
Model~5 predicts $kT_{\rm e}(r)\simeq0.5~{\rm keV}\simeq E_{u,l}$ therefore the predicted O\emissiontype{VII}~He$\alpha$ line would be the
brightest among the models.
\par
Figure~\ref{fig:oxygen abundance rcw86} shows the radial profile of the oxygen abundance $n_{Z,z}/n_Z$
for Model~2 (left panel) and Model~5 (right panel). The O\emissiontype{VII} abundance (orange) is large. Note that
the other models (e.g., Model~0) also result in a large O\emissiontype{VII} abundance.
Model~2 predicts the smaller abundance of O\emissiontype{VII} than the case of Model~5 because
the higher electron temperature results in a faster ionization.
The temperature of O\emissiontype{VII} is approximately $kT_{Z,z}(r)\approx 250~{\rm keV}\times\left[r/R_{\rm sh}(t_{\rm age})\right]$ for Model~2 and
$kT_{Z,z}(r)\approx 140~{\rm keV}\times\left[r/R_{\rm sh}(t_{\rm age})\right]$ for Model~5.
\par
We estimate the line emission as follows: the observed specific intensity per frequency $I_\nu$ at the sky position
${\cal X}$ from the center of the SNR is calculated as
%
%%%%%%%%%%%%%
\begin{eqnarray}
I_\nu({\cal X}) = \int_{-L}^{L}
\left\{
\int_{-\infty}^{\infty}{\cal J}_\nu {\cal G}\left(w_t,v_{\cal Z}\right) dw_t \right\}
d{\cal Z},~~~
\end{eqnarray}
%%%%%%%%%%%%%
%
where $L=\sqrt{R_{\rm sh}-{\cal X}^2}$.
The position along the line of sight is ${\cal Z}$ so that $r=\sqrt{{\cal X}^2+{\cal Z}^2}$.
$v_{\cal Z}(r)=({\cal Z}/r)v(r)$ is the line of sight velocity.
The probability distribution function of the turbulence ${\cal G}$ is assumed to be a Gaussian as
%
%%%%%%%%%%%%%
\begin{eqnarray}
{\cal G}(w_t,v_{\cal Z})
=\frac{1}{\sqrt{\pi}v_{\rm turb}}
\exp\left[-\frac{ m_Z\left( w_t - v_{\cal Z} \right)^2 }{ 2K_Z } \right],
\end{eqnarray}
%%%%%%%%%%%%%
%
where $v_{\rm turb}$ is a typical turbulent velocity and $K_Z\equiv(1/2)m_Z v_{\rm turb}{}^2$. Note that $w_t$ is the variable
for the integration. In this paper, we assume that the intensity of the turbulence is proportional to the proton sound speed as $v_{\rm turb}(r)=\delta
\sqrt{\gamma kT_{\rm p}(r)/m_{\rm p}}$. Supposing the incompressible turbulence is driven in the downstream region~(see \cite{shimoda18b}), we
calculate the case of $\delta=0.5$ and the case without the turbulent Doppler broadening $\delta=0$ for a comparison.
The emissivity of the line is given by
%
%%%%%%%%%%%%%
\begin{eqnarray}
{\cal J}_\nu = \frac{W_{Z,z}^{u,l}}{4\pi} n_{\rm e} n_{Z,z} \phi_\nu,
\end{eqnarray}
%%%%%%%%%%%%%
%
where we have neglected the cascade from the higher excitation levels. The line profile function is defined as
%
%%%%%%%%%%%%%
\begin{eqnarray}
&& \phi_\nu\equiv\frac{1}{\sqrt{\pi}\Delta\nu}\exp\left[-\left(\frac{\nu-\nu_0}{\Delta\nu}\right)^2\right], \\
&& \Delta\nu = \nu_0'\sqrt{ \frac{2kT_{Z,z}}{m_Zc^2} }, \\
&& \nu_0     = \nu_0'\left( 1 + \frac{v_{\cal Z}}{c} \right),
\end{eqnarray}
%%%%%%%%%%%%%
%
where $\nu_0'$ is the frequency of the line measured in the rest frame of the atom.
Then, we obtain
%
%%%%%%%%%%%%%
\begin{eqnarray}
I_\nu({\cal X})
&=& \int_{-L}^{L}
\frac{n_{\rm e}n_{Z,z}W_{Z,z}^{u,l}}{4\pi\Delta\nu\sqrt{\pi\left(1+{\cal M}_{Z,z}{}^2\right)}}
\nonumber \\
&\times&
\exp\left[-
\left\{
\frac{ \nu - \nu_0'\left(1+v_{\cal Z}/c\right) }{ \Delta\nu\sqrt{1+{\cal M}_{Z,z}{}^2} }
\right\}^2\right] d{\cal Z},
\end{eqnarray}
%%%%%%%%%%%%%
%
where ${\cal M}_{Z,z}{}^2\equiv K_Z/kT_{Z,z} = (\gamma\delta^2/2)(m_Z/m_{\rm p})(T_{\rm p}/T_{Z,z})$.
The line shape is broadened by the bulk Doppler effect $(1+v_{\cal Z}/c$) and the turbulent Doppler effect
$\sqrt{1+{\cal M}_{Z,z}{}^2}$.
\par
%
%%%%%%%%%%%%%
\begin{figure}
\begin{center}
\includegraphics[scale=0.65]{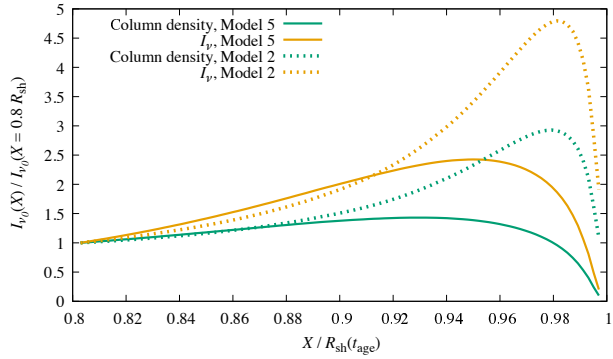}
\end{center}
\caption{
The radial profile of the specific intensity $I_\nu({\cal X})/I_\nu(0.8R_{\rm sh})$ at the line center for
O\emissiontype{VII}~He$\alpha$ (orange). The solid lines show the results of Model~5 (multiplied by a factor of 10) and
the dots show Model~2. We also display profiles of the column density (green) of O\emissiontype{VII}.}
\label{fig:projected rcw86}
\end{figure}
%%%%%%%%%%%%%
%
Figure~\ref{fig:projected rcw86} shows $I_\nu({\cal X})/I_\nu(0.8R_{\rm sh})$ at the line center for Model~5 (solid lines)
and Model~2 (dots). We also display profiles of the column density of
O\emissiontype{VII} (green). The difference between the column density profile and the intensity profile results from the excitation.
The spatial variation of the 
electron temperature is relatively less important in this case because the excitation rate depends on $\exp(-E_{ul}/kT_{\rm e})/\sqrt{T_{\rm e}}$
that is not so sensitive on $T_{\rm e}$ unless $kT_{\rm e}\ll E_{ul}$.
\par
%
%%%%%%%%%%%%%
\begin{figure}
\begin{center}
\includegraphics[scale=0.65]{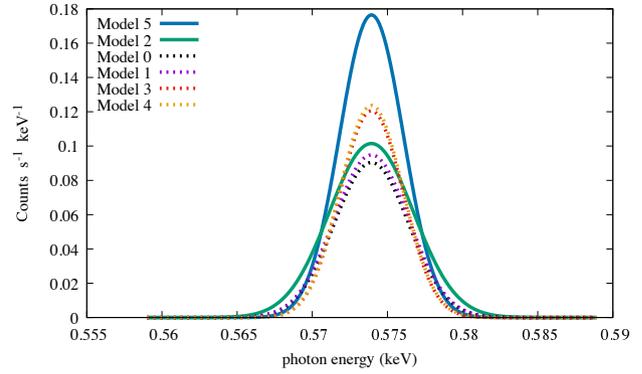}
\end{center}
\caption{The calculated O\emissiontype{VII}~He$\alpha$ line with $\delta=0.5$ for Model~5 (blue solid line) and Model~2 (green
solid line). We also display the results of Model~0 (black dots), Model~1 (purple dots), Model~3
(red dots), and Model~4 (orange dots). We assume the distance of the RCW~86 is $2.5$~kpc and the
observed area is $0.2R_{\rm sh}\times0.2R_{\rm sh}$, where $R_{\rm sh}=15.27~{\rm pc}$.}
\label{fig:line rcw86}
\end{figure}
%%%%%%%%%%%%%
%
Figure~\ref{fig:line rcw86} shows the calculated O\emissiontype{VII}~He$\alpha$ line for Model~5 (blue solid line) and Model~2 (green
solid line) derived from $\int I_\nu d{\cal X}$ with $\delta=0.5$.
We assume the distance of the RCW~86 as $2.5$~kpc~\citep{yamaguchi16} and the
observed area as $0.2R_{\rm sh}\times0.2R_{\rm sh}$, where $R_{\rm sh}=15.27~{\rm pc}$.
We also display the results of Model~0 (black dots), Model~1 (purple dots), Model~3
(red dots), and Model~4 (orange dots).  The results show a good agreement
with the observed photon counts $\sim0.15~{\rm counts~s^{-1}~keV^{-1}}$~\citep{vink06}.
Table~\ref{tab:line summary} shows a summary of the calculated O\emissiontype{VII}~He$\alpha$ line.
The derived temperatures reflect the effects of the efficient CR acceleration.
From the comparison of $\delta=0.5$ to $\delta=0$, the turbulent Doppler broadening results in
the higher observed temperatures by a factor of $\sim1.05$. The degree of the broadening can be estimated as
$\sqrt{1+{\cal M}_{Z,z}{}^2}\approx1.1$ for $\delta=0.5$ with approximating $T_{\rm p}/T_{Z,z}\approx m_{\rm p}/m_Z$.
Since the observed line consists of multiple temperature populations, and since a higher temperature population less contributes
around the line center, a lower temperature population is accentuated around the line center.
The contribution of the higher temperature population appears far from the
line center like a `wing'. If we measure the temperature using the full width at the e-folding scale, the difference in
the derived temperatures becomes large. Hence the observed FWHM is smaller than that expected from $\sqrt{1+{\cal M}_{Z,z}{}^2}$.
\par
The RCW~86 also shows bright Ne\emissiontype{IX}~He$\alpha$ however, our model predicts a faint Ne\emissiontype{IX}~He$\alpha$
emission (the intensity is smaller than a tenth of O\emissiontype{VII}~He$\alpha$ intensity).
The line intensity also depends on the ion abundance. In this paper, we use the solar abundance that reflects the condition
of our galaxy $\simeq4.6~{\rm Gyr}$ ago.
Moreover, \citet{decia21} found large variations of the chemical abundance of the neutral ISM
in the vicinity of the Sun over a factor of 10 (Si, Ti, Cr, Fe, Ni, and Zn they analyzed).
Their findings imply that the gaseous matter is not well mixed.
The predicted faint Ne\emissiontype{IX}-He$\alpha$ might reflect
a different abundance pattern from the solar abundance pattern.
%
%%%%%%%%%%%%%%
\begin{table}[htbp]
\tbl{Summary of the calculated O\emissiontype{VII}~He$\alpha$. From the left-hand side to the right-hand side,
the columns indicate the model name, the O\emissiontype{VII} temperature derived from the FWHM of the line for the case
of $\delta=0.5$, and the O\emissiontype{VII} temperature derived from the FWHM of the line for the case
of $\delta=0$.}{%
\begin{tabular}{c cc}
\hline
Model & $kT_{Z,z}$ ($kT_{Z,z}/2Z$) for $\delta=0.5$ &  for $\delta=0$                              \\
0     & 325.9~keV (20.4~keV)                        &  312.3~keV (19.5~keV)                        \\
1     & 325.6~keV (20.3~keV)                        &  312.4~keV (19.5~keV)                        \\
2     & 306.4~keV (19.2~keV)                        &  296.5~keV (18.5~keV)                        \\
3     & 160.6~keV (10.0~keV)                        &  153.8~keV (9.62~keV)                        \\
4     & 160.6~keV (10.0~keV)                        &  154.2~keV (9.63~keV)                        \\
5     & 157.7~keV (9.85~keV)                        &  152.5~keV (9.53~keV)                        \\
\hline                                
\end{tabular}}                        
\label{tab:line summary}
\end{table}
%%%%%%%%%%%%%%
%
\par
%
%%%%%%%%%%%%%
\begin{figure}
\begin{center}
\includegraphics[scale=0.65]{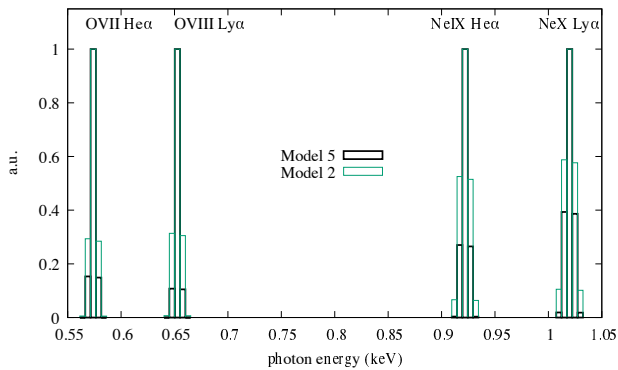}
\end{center}
\caption{The line profiles with $5$~eV resolution for $\delta=0.5$.
We display O\emissiontype{VII}~He$\alpha$, O\emissiontype{VIII}~Ly$\alpha$, Ne\emissiontype{IX}~He$\alpha$, and
Ne\emissiontype{X}~Ly$\alpha$
for Model~5 (black solid line) and Model~2 (green solid line).}
\label{fig:line bin rcw86}
\end{figure}
%%%%%%%%%%%%%
%
Figure~\ref{fig:line bin rcw86} represents the line shape with $5$~eV resolution for $\delta=0.5$.
We additionally show O\emissiontype{VII}~Ly$\alpha$, Ne\emissiontype{IX}~He$\alpha$, and Ne\emissiontype{X}~Ly$\alpha$. Since the widths
of the particle distribution function are almost the same as each other for $nt\sim10^9\mathchar`-10^{11}~{\rm cm^{-3}~s}$,
the observation of lines at higher photon energy is better to resolve the line width.
Note that the observed O\emissiontype{VII}~He$\alpha$  and
Ne\emissiontype{IX}~He$\alpha$ are bright compared to the continuum emission~\citep{vink06}.
The energy resolution of {\it XRISM}'s micro-calorimeter
{\it Resolve} is sufficient to distinguish whether the SNR shock accelerates the CRs (Model~5) or not (Model~2).

\section{Summary and Discussion}
\label{sec:discussion}
We suggest the novel collisionless shock jump condition, which is given by modeling each ion species' entropy production
at the shock transition region. As a result, the amount of the downstream thermal energy is given.
The magnetic-field amplification driven by the CRs is assumed.
For the given strength of the amplified field, the amount of the CRs is constrained by the energy conservation law.
The constrained amount of the CRs can be sufficiently large to explain the Galactic CRs.
The ion temperature is lower than the case without the CRs because the upstream kinetic energy
is divided into the CRs and the amplified field.
The strength of the filed around the shock transition region is assumed to be
$1/\sqrt{\xi_{\rm B}}=v_0/(\delta B/\sqrt{4\pi\rho_0})\simeq 3$.
Downstream developments of the ionization balance and temperature relaxation are also calculated. Using
the calculated downstream values, we perform
synthetic observations of atomic lines for the SNR RCW~86, including the Doppler broadening by the turbulence.
Our model predictions can be consistent with the previous observations of the SNR RCW~86, and the predicted line widths
are sufficiently broad to be resolved by the {\it XRISM}'s micro-calorimeter.
Future observations of the X-ray lines
can distinguish whether the SNR shock accelerates the CRs or not from the ion temperatures.
\par
Our shock model constrains the maximum fraction of the CRs depending on the shock velocity, the upstream density,
and the sonic Mach number (see figure~\ref{fig:Ma vs csi}). Since the SNR shock decelerates gradually, we can predict
a history of the CR injection and related nonthermal emissions, especially the hadronic $\gamma$-ray emissions.
Although the injection history of the CRs is essential to estimate the intrinsic injection
of the CRs into our galaxy per one supernova explosion, this issue currently remains to be resolved~(e.g., \cite{ohira10,ohira11}).
The injected CRs will contribute to the dynamics of the ISM as a pressure source, leading to
a feedback effect on the star formation rate, for example~(e.g., \cite{hopkins18,girichidis18,shimoda21}).
The origin of $\gamma$-ray emissions in the SNRs
is also unsettled, whether the hadronic origin or leptonic origin~(\cite{abdo11}, but see \cite{fukui21}).
We will study them in a forthcoming paper.
\par
For distinguishing the case of extremely efficient CR acceleration (Model~3) from the case of no CRs (Model~0),
a comparison of the FWHM to other values  is required in general (e.g., the difference between the ionization states,
the shock velocity, and so on). The FWHM of Model~3 becomes smaller than
Model~0 at a given shock velocity and $nt$, and abundant ions of Model~3 tend to be less ionized than
in the case of Model~0 because of the lower electron temperature.
The lower electron temperature and lower ionization states of Model~3 may result in a different photon spectrum from the case
of other Models, especially the equivalent widths, recombination lines, Aug{\'e}r transitions
due to the inner shell ionization, and so on. We will attempt further investigations by calculating
the overall photon spectrum in future work.
\par
The line diagnostics of the thermal plasma of young SNRs on the effect of CR acceleration will be a good science
objective for the {\it XRISM} mission \citep{xrism20}, which will provide high-resolution X-ray spectroscopy.
Since the micro-calorimeter array is not a distributed-type spectroscope like grating optics on {\it Chandra} and/or
{\it XMM-Newton}, the {\it Resolve} onboard {\it XRISM} \citep{ishisaki18} can accurately measure the atomic-line profiles in
the X-ray spectra from diffuse objects like SNRs. The {\it XRISM} will have the energy resolution of 7 eV (as the design
goal), and the calibration goals on the energy scale and resolution are 2 eV and 1 eV, respectively \citep{miller20}.
Therefore, the line broadening values from multiple elements with/without CRs in figure~\ref{fig:initial} can be distinguished by {\it XRISM}.
Another importance of {\it XRISM} is the wider energy coverage, with which atomic lines not only from light elements (C, N, O, etc.)
but also Fe will be measured. So, the intensity of the turbulence demonstrated in section \ref{sec:synthetic} will be constrained
with {\it XRISM}. The preparation for the instruments \citep{nakajima20,frederick20} and in-orbit operations \citep{terada21,
loewenstein20} are proceeding smoothly for the launch in 2022/2023, and several young SNRs, including RCW86 are
listed as the target during the performance verification phase of {\it XRISM}.
\footnote{$\langle$https://xrism.isas.jaxa.jp/research/proposer/approved/pv/index.html$\rangle$}
We expect to verify our predictions observationally soon.

\begin{ack}
We thank K. Masai and G. Rigon for useful discussions.
We are grateful to the anonymous referee, for his/her comments that further improved the paper.
This work is partly supported by JSPS Grants-in-Aid for Scientific Research Nos. 20J01086 (JS),
19H01893 (YO),
JP21H04487 (YO)
19K03908 (AB),
20K04009 (YT), 
18H01232 (RY),
22H01251 (RY),
and 20H01944(TI).
YO is supported by Leading Initiative for Excellent Young Researchers, MEXT, Japan.
RY and SJT deeply appreciate Aoyama Gakuin University Research Institute for
helping our research by the fund.
\end{ack}

\appendix 
%\section*{Case of single paragraph}
%\section{Case of two or more paragraphs}
%\section{Case of two or more paragraphs}

%%%
% See the manual for the detail.
%%%
%\begin{thebibliography}{}
% Journals(e.g. A\&A,ApJ,AJ,NMRAS,PASP ...)
% Authors, Year, Journal, Vol#, Page#
% Journal Title Abbreviation >> http://www.asj.or.jp/pasj/Jabb.html
%\bibitem[Aauthor et al.(2001)]{key-1}
%  Aauthor, A., Bauthor, B., Cauthor, C.\ 2001, PASJ, vol, page
%\bibitem[Aauthor \& Bauthor(2003a)]{key-2}
%  Aauthor, A., \& Bauthor, B.\ 2003a, PASJ, vol, page   
%\bibitem[Aauthor \& Bauthor(2003b)]{key-3}
%  Aauthor, A., \& Bauthor, B.\ 2003b, PASJ, vol, page  
%\bibitem[Aauthor, Cauthor, and Dauthor(2000)]{key-3}
%  Aauthor, A., Cauthor, C., \& Dauthor, D.\ 2000, PASJ, vol, page   
%% Books
%\bibitem[Aauthor \& Eauthor(2003b)]{key-3}
%  Aauthor, A., \& Euthor, E.\ 2003b, Name of Book (Tokyo: Publisher) ch.0    
%% Editorial Books
%\bibitem[Dauthor(2001)]{key-n}
%  Dauthor A.~A.\ 2001, in Name of Book,
%   ed.\  D.~Editor (Tokyo: Publisher), page
%\end{thebibliography}

\bibliography{apj_sjsi21}{}
\bibliographystyle{apj.bst}

\end{document}